\documentclass{acmsiggraph}                     

\usepackage{booktabs} 
\usepackage{microtype}
\usepackage{amsmath}
\usepackage{amssymb}
\usepackage{units}
\DeclareMathAlphabet{\mathcal}{OMS}{cmsy}{m}{n}  
\usepackage{xfrac}
\usepackage[normalem]{ulem}
\usepackage{graphicx}  		
\usepackage{float}
\usepackage{url}
\usepackage{xspace}
\usepackage{color}
\usepackage[normalem]{ulem}  
\usepackage{enumitem}  		
\usepackage{tikz}			
\usepackage{calc}			
\usepackage{collcell}
\usepackage{tabularx}

\usepackage{multirow}
\usepackage{amsmath}
\usepackage{amssymb}
\usepackage{amsfonts}
\usepackage{eucal}
\usepackage[latin1]{inputenc}
\usepackage[normalem]{ulem}
\usepackage{listings}
\usepackage{mathrsfs}
\usepackage{subcaption}
\usepackage{cancel}
\usepackage[normalem]{ulem}
\usepackage{cancel}
\usepackage{tcolorbox}
\usepackage{verbatimbox}

\usepackage{graphicx}
\usepackage{color}
\usepackage{wrapfig}
\usepackage{ifthen}
\usepackage[]{algorithm2e}

\title{Charted Metropolis Light Transport}

\author{Jacopo Pantaleoni\thanks{e-mail: jpantaleoni@nvidia.com}\\NVIDIA}


\begin{document}
	
	\newcommand{\picresdir}{final}

	\teaser{
		\includegraphics[width=180.0mm]{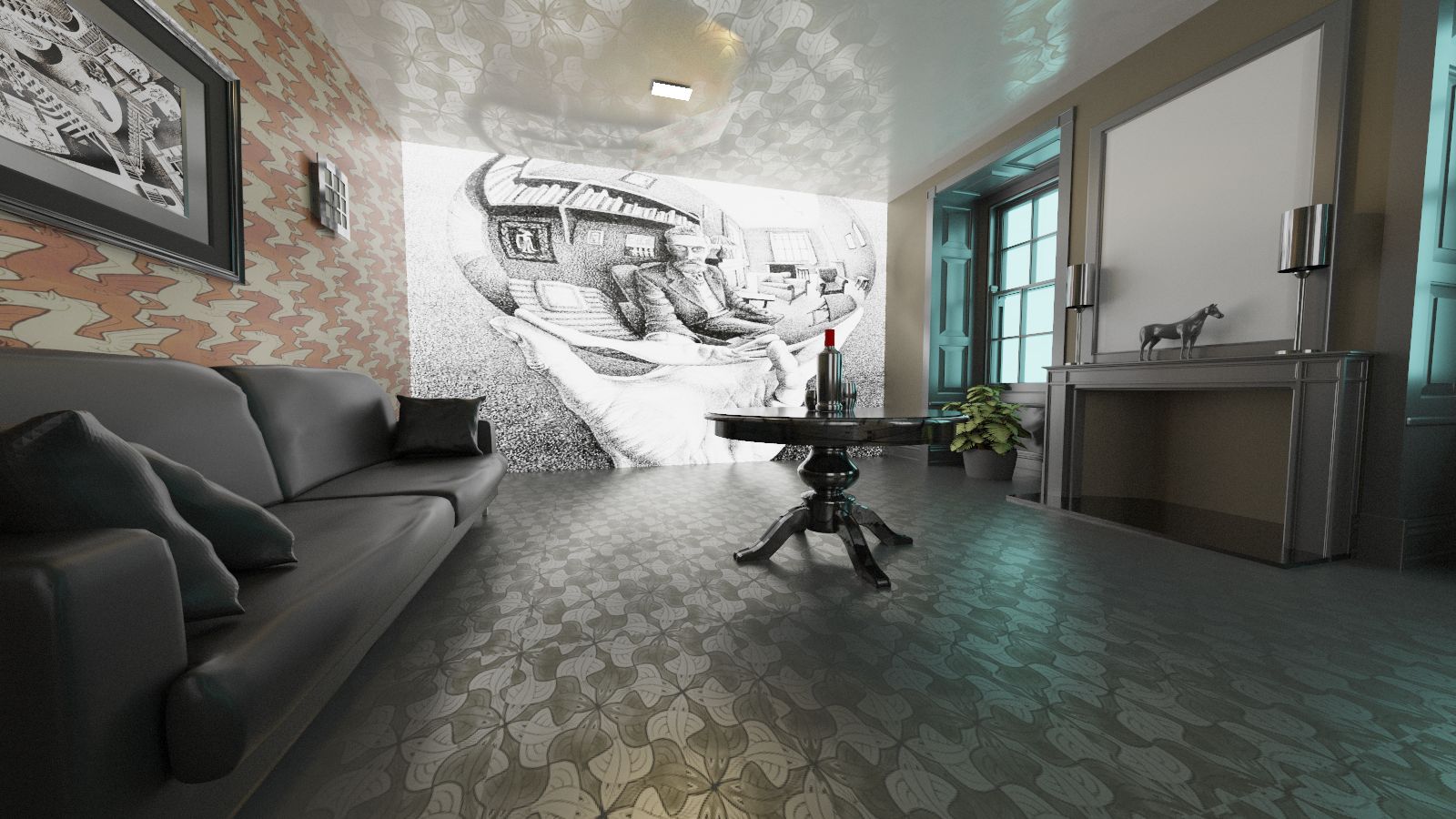}
		\caption{
			\emph{Escher's Room.}
			Charted Metropolis light transport considers path sampling methods and their primary sample space coordinates as charts of the path space, allowing to easily jump between them. In particular, it does so without requiring classical invertibility of the sampling methods, making the algorithm practical even with complex materials.
		}
	}
	
	
	\maketitle
	
	
	\begin{abstract}
		
		In this manuscript, inspired by a simpler reformulation of primary sample space Metropolis light transport, we derive a novel family of general Markov chain Monte Carlo algorithms called \emph{charted Metropolis-Hastings}, that introduces the notion of \emph{sampling charts} to extend a given sampling domain and make it easier to sample the desired target distribution and escape from local maxima through coordinate changes.
		We further apply the novel algorithms to light transport simulation, obtaining a new type of algorithm called \emph{charted Metropolis light transport}, that can be seen as a bridge between primary sample space and path space Metropolis light transport.
		The new algorithms require to provide only right inverses of the sampling functions, a property that we believe crucial to make them practical in the context of light transport simulation.
		We further propose a method to integrate density estimation into this framework through a novel scheme that uses it as an independence sampler. 
		
	\end{abstract}
	
	
	\begin{CRcatlist}
		\CRcat{I.3.2}{Graphics Systems C.2.1, C.2.4, C.3)}{Stand-alone systems};
		\CRcat{I.3.7}{Three-Dimensional Graphics and Realism}{Color,shading,shadowing, and texture}{Raytracing};
	\end{CRcatlist}

	\keywordlist
	
	
	\copyrightspace

\section{Introduction}

Light transport simulation can be notoriously hard. The main problem is that forming an image requires evaluating millions of infinite dimensional integrals, whose integrands, while correlated, may contain an infinity of singularities and different modes at disparate frequencies.
Many approaches have been proposed to solve the rendering equation, though most of them rely on variants of Monte Carlo integration.
One of the most robust algorithms, Metropolis light transport (MLT), has been proposed by Veach and Guibas in 1997 \cite{Veach:1997:MLT} and has been later extended in many different ways.
One of the most commonly used variants is primary sample space MLT \cite{Kelemen:2002}, partly because in some scenarios it is more efficient (though not always), partly because it is generally considered simpler to implement.
However, both variants are still considered relatively complex compared to other algorithms that are not based on Markov chain Monte Carlo (MCMC) methods, or that employ a simplified target distribution \cite{Hachisuka:2011}.

In this paper we show that the original primary sample space MLT uses a suboptimal target distribution, and that fixing the problem makes the algorithm more efficient while also greatly simplifying it at the same time.

Inspired by this simpler formulation, we then propose a novel family of general Markov chain Monte Carlo algorithms called \emph{charted Metropolis-Hastings} (CMH).
The core idea is to extend the concept of primary sample spaces into that of \emph{sampling charts} of the target space, extending the domain of the desired target distribution and introducing novel mutation types that swap charts and perform coordinate changes (analogous to those found in regular tensor calculus) in order to craft better proposals.

We then apply the new MCMC algorithm to light transport simulation, obtaining a type of algorithms called \emph{charted Metropolis light transport} (CMLT), that considers all local path sampling methods as parameterizations of the path space manifold, and employs stochastic path inversion as a way to perform coordinate transformations between charts.
Our algorithm is made practical by avoiding the requirement to use fully invertible path sampling methods - a property we believe fundamental - and only requiring stochastic right inverses.
This new type of algorithms can be seen as fundamentally bridging the difference between the original formulation of path space MLT and the primary sample space version, allowing to easily combine both.

Finally, we briefly propose a novel scheme to integrate density estimation inside MCMC frameworks that exploits its robustness with respect to sampling near-singular and singular paths while mantaining overall simplicity and efficiency of implementation.


\section{Main contribution}

The main contribution of our paper is extending primary sample space MLT \cite{Kelemen:2002} by introducing mutations that allow to \emph{swap} bidirectional sampling techniques at any time \emph{while preserving the underlying path}.
Alternatively, adopting a different viewpoint, we could say our main contribution is allowing to freely apply all types of primary space mutations to any given path.

The key strength, missing from the original primary space formulation, is allowing to break the path in the middle at any arbitrary point along it and mutate the two resulting subpaths using the corresponding primary space perturbations, bringing back the flexibility of path space MLT, combined with primary space BSDF importance sampling.

This is achieved in two ways: the first is realizing that the single primary sample space defined in the original work of Kelemen et al \shortcite{Kelemen:2002} can be more flexibly thought of as a \emph{collection} of different primary sample spaces stitched together through Russian Roulette, with each space corresponding to a specific bidirectional sampling technique.

The second is realizing that each primary sample space is nothing more than a parameterization of path space, and that if we could \emph{invert} them we could effectively transform this set of parameterizations into a proper atlas, where each primary space is a chart.
Once this is achieved, crafting mutations that jump between the charts while not changing the represented path is just a matter of applying proper transformations and following the rules for mantaining detailed balance.

However, this second step is made complicated by the fact that the parameterizations typically used in bidirectional path tracers are not always classically invertible, making it impossible to unambiguosly recover the primary space coordinates of a given path. In fact, in the presence of layered materials, sampling the BSDF, which is at the core of any local path sampling technique, is often based on the use of non-injective maps from primary coordinates to the sphere of outgoing directions: for example, if a diffuse and a glossy layer are present, each outgoing direction might be sampled by both layers. In these cases the local primary sample space corresponding to each scattering event is typically divided in two or more strata, each of which maps to the entire sphere (or hemisphere) of directions.

As this means we cannot employ the notion of charts used in standard manifold geometry, which requires the parameterizations to be invertible, we hence introduce the notion of \emph{sampling charts}, that unlike the deterministic counterpart doesn't rely on classical inverses, but rather requires to only provide stochastic right inverses.
This new definition allows to move freely between different primary sample spaces even in cases of ambiguity, employing the probability densities associated to these stochastic inverses to compute the transition probabilities needed to satisfy detailed balance.

The rest of the paper is dedicated to explaining our framework in detail.
In particular, the following sections are organized as follows:
section 3 introduces some preliminaries required to properly frame the problem, as well as a simpler reformulation of primary sample space MLT in which all the primary spaces are kept explicitly separate; section 4 introduces our new framework in a very abstract and general mathematical setting; finally section 5 details its application to light transport simulation, and section 6 and 7 are dedicated to describing our massively parallel implementation of the algorithms, and providing test results.

This paper is a preprint of a SIGGRAPH publication \cite{SelfSiggraph:2017}.
Concurrent to our work Otsu et al \shortcite{Anon:0462} have developed 
a novel set of mutations relying on an inverse mapping from path space to primary sample space:
while proposing different solutions and mathematical methods, our algorithms share a similar underlying idea.

\section{Preliminaries}

Veach \shortcite{Veach:PHD} showed that light transport simulation can be expressed as the solution of per-pixel integrals of the form:
\begin{equation}
I_j = \int_{\Omega} f_j({\bf x}) d\mu({\bf x})
\end{equation}
where $\Omega = \bigcup_{k=1}^{\infty} \Omega_k$ represents the space of light paths of all finite lengths $k$ and $\mu$ is the area measure, and $j$ is the pixel index.

For a path ${\bf x} = x_0 \rightarrow x_1 \dots \rightarrow x_k$, the integrand is defined by the \emph{measurement contribution function}:
\begin{eqnarray}
f_j({\bf x}) &=&
L_e(x_0 \rightarrow x_1) \nonumber \\
&\cdot& \prod_{i=0}^{k-1} \big[ f_s(x_{i-1} \rightarrow x_i \rightarrow x_{i+1}) G(x_i \leftrightarrow  x_{i+1}) \big] \nonumber \\
&\cdot& W_e^j(x_{k-1} \rightarrow x_{k})
\end{eqnarray}

where $L_e$ is the surface emission, $W_e^j$ is the pixel response (or emitted importance), $f_s$ denotes the local BSDF and $G$ is the geometric term.
To simplify notation, in the following we will simply omit the pixel index and consider the positions $f = f_j$ and $I = I_j$.

Veach further showed that if one employs a family $\mathcal{F}_k = \{s,t : s+t-1 = k\}$ of \emph{local path sampling} techniques to sample subpaths ${\bf y} = {y_0 \dots y_{s-1}}$ and ${\bf z} = {z_0 \dots z_{t-1}}$ from the light and the eye respectively, and build the joined path ${\bf x} = y_0 \dots y_{s-1} z_{t-1} \dots z_0$, an unbiased estimator of $I$ can be obtained as a \emph{multiple importance sampling} combination:
\begin{equation}
F = \sum_{s,t} C_{s,t}({\bf x})
\end{equation}
with the following definitions:
\begin{equation}
C_{s,t}({\bf x}) = w_{s,t}C^*_{s,t}
\end{equation}
\begin{equation}
C^*_{s,t}({\bf x}) = \frac{f({\bf x})} {p_{s,t}({\bf x})} 
\end{equation}
\begin{equation}
p_{s,t}({\bf x}) = p_s({\bf x}) p_t({\bf x})
\end{equation}
\begin{equation}
w_{s,t} = \frac{p_{s,t}({\bf x})} { \sum_{(i,j) \in \mathcal{F}_k} p_{i,j}({\bf x})}
\end{equation}

While a complete analysis of the above formulas is beyond the scope of this paper (we refer the reader to \cite{Veach:PHD}), we feel it is important to make the following:

\paragraph{Remark:} if importance sampling is used, the connection term $C^*_{s,t}$ effectively contains only the parts of $f$ which have not been importance sampled; particularly, if $p_s$ and $p_t$ importance sample all terms of the measurement contribution function up to the $s$-th and $t$-th light and eye vertex respectively, $C^*_{s,t}$ will be proportional to the BSDFs at the connecting vertices times the geometric term $G(y_{s-1},z_{t-1})$. This is the only remaining singularity, which gets eventually suppressed in $C_{s,t}$ by the multiple importance sampling weight $w_{s,t}$. In fact, simplifying equation (4), one gets:
\begin{equation}
C_{s,t}({\bf x}) = \frac{f({\bf x})}{ \sum_{(i,j) \in \mathcal{F}_k} p_{i,j}({\bf x}) } \nonumber
\end{equation}

\subsection{The Metropolis-Hastings Algorithm}

The Metropolis-Hastings algorithm is a Markov-Chain Monte Carlo method that, given an arbitrary target distribution $\pi(x)$, builds a chain of samples $X_1, X_2, \dots $ that have $\pi$ as the stationary distribution, i.e. $\lim_{n \rightarrow \infty} p(X_n) = \pi(X_n)$.
The algorithm is based on two simple steps:

\paragraph{proposal:} a new sample $Y$ is obtained from $X=X_i$ by means of a \emph{transition kernel} $K(Y|X)$

\paragraph{acceptance-rejection:}
$X_{i+1}$ is set to $Y$ with probability:
\begin{equation}
A(Y|X) = \min \left( 1, \frac{\pi(Y)K(X|Y)}{\pi(X)K(Y|X)} \right)
\end{equation}
and to $X_i$ otherwise.

Importantly, note that $\pi$ can be defined up to a constant. In other words, if $\int \pi(x) dx = c$, the algorithm will simply admit $\pi/c$ as its stationary distribution.

Finally, it is also possible to use mutations in which the proposal $K(Y|X) = K(Y)$ depends only on $Y$: in this case, the mutation type is called an independence sampler \cite{Tierney:1994}.

\subsection{Primary sample space Metropolis light transport, revisited}

Kelemen et al \shortcite{Kelemen:2002} showed that if one considers the transformation $T : U \rightarrow \Omega$ that is typically used to map random numbers to paths when performing forward and backward path tracing (i.e. when sampling eye and light subpaths), one can
apply the Metropolis-Hastings algorithm on the unit hypercube $U$ instead of working in the more complex path space.
The advantage is that crafting mutations in $U$ is much easier to implement - a simple Gaussian kernel will do - and will often lead to better mutations, since they will naturally follow the local BSDFs.\footnote{This can, however, be detrimental in cases of complex occlusion, where the original path space MLT is generally superior. The reason is that the BSDF parameterizations might squeeze unoccluded, off-specular directions into vanishingly small regions of the primary sample space.}
The only requirement is pulling back the desired measure from $\Omega$ to $U$, which is easily achieved by multiplying by the Jacobian of the transformation $T$, which is nothing more than the reciprocal of path probability:
\begin{equation}
I = \int_U f(T(u)) \left|\frac{dT(u)}{du}\right| du = \int_U \frac{ f(T(u)) } {p(T(u))} du
\end{equation}

We now provide a novel formulation that improves the choice of mapping and target distributions compared to the ones employed by Kelemen et al \shortcite{Kelemen:2002}.


In fact, what was done in the original work was to consider a mapping from the product of two infinite-dimensional unit hypercubes,\footnote{A formulation which, technically, poses some definition challenges, as infinite dimensional spaces do not possess a Lebesgue measure.} to the product space of light and eye subpaths sampled using Russian Roulette terminated path tracing.
Furthermore, instead of simply considering the single path obtained by joining the two endpoints of the respective subpaths, and using the measurement contribution function as the target distribution, they considered the sum of the MIS weighted contributions from all paths obtained joining any two vertices of the light and eye subpaths.
The reason why this was done can be understood: this was the historical way to perform bidirectional path tracing. In order not to \emph{waste} any vertex, one would \emph{reuse} all of them at the expense of some added correlation and some added shadow rays.
However, this is undesirable for several reasons:

\paragraph{1.} by joining all vertices in the generated subpaths, and summing up all the weighted contributions from the obtained paths (which are in fact truly different paths, except for the fact they share their light and eye prefixes), they were using \emph{a target distribution which was no longer proportional to path throughput} (or, more precisely, the measurement contribution function we are finally interested in).
In other words, the obtained paths have a \emph{skewed} distribution which is not necessarily optimal.\footnote{One can consider their technique to generate \emph{path bundles} and in this sense their target distribution is optimal for the constructed bundles, relative to the overall bundle contribution, but not for the individual paths.}

\paragraph{2.} dealing with the infinite dimensional unit hypercubes introduces some unnecessary algorithmic complications, including the need for lazy coordinate evaluations.

\paragraph{3.} by joining all vertices in the generated subpaths, we are introducing some additional sample correlation that might not necessarily improve the per-sample efficiency. In some situations, for example in the presence of incoherent transport or complex occlusion, it will in fact reduce it.

\vspace{2mm}
In light of these problems, we now propose a much simpler variant.
Let's for the moment consider the space of paths of length $k$, and a single technique $i \in \mathcal{F}_k$ to generate them, where $i$ defines the number of light vertices and the number of eye vertices is given as $j = k+1-i$.
If sampling $n$ vertices through path tracing requires $m \cdot n$ random numbers, we will consider the following definition of the primary sample space:
\begin{equation}
U_i = [0,1]^{m \cdot i} \times [0,1]^{m \cdot (k+1-i)}.
\end{equation}
The transformation $T = T_i: U \rightarrow \Omega_k$ will have the following Jacobian:
\begin{equation}
\left| \frac{dT(u)}{du} \right| = \frac{1}{ p_{i}(T(u)) }.
\end{equation}
We now have two options for the choice of our target distribution.
The simplest is to set:

\vspace{2mm}
{\bf Definition}: \emph{Importance sampled distributions}
\begin{equation}
\pi_i(u) = \frac{ f(T(u)) }{ p_{i}(T(u)) }.
\label{eqn:ISDistributions}
\end{equation}
This choice keeps the corresponding path space distribution invariant relative to the area measure $\mu$, as we have:
\begin{eqnarray}
\pi_i(u) du
&=& \pi_i(u) p_i(T(u)) |d\mu(T(u))/du|du \nonumber \\
&=& \pi_i(u) p_i(T(u)) d\mu(T(u)) \nonumber \\
&=& f(T(u)) d\mu(T(u)) \nonumber \\
&=& \bar{\pi}(T(u)) d\mu(T(u)).
\end{eqnarray}
In other words, it ensures that all our distributions $\pi_i(u)$ are designed to have a distribution in their primary space $U_i$ that becomes the same distribution $\bar{\pi}({\bf x}) = f({\bf x})$ in path space.

\noindent The second choice is to use the following:

\vspace{2mm}
{\bf Definition}: \emph{Weighted distributions}
\begin{equation}
\pi_i(u) = w_i(T(u)) \frac{ f(T(u)) }{ p_i(T(u)) },
\label{eqn:WDistributions-1}
\end{equation}
exploiting the fact that, while now the corresponding path space distributions
$\bar{\pi}_i({\bf x}) = w_i({\bf x})f({\bf x})$ are biased,\footnote{In practice instead of sampling $f$, they are sampling a version downscaled locally according to the efficiency of $p_i$} our desired path space distribution $f$ is obtained as their sum:
\begin{equation}
\sum_{i\in\mathcal{F}_k}\bar{\pi}_i({\bf x})
= \sum_{i\in\mathcal{F}_k} w_i({\bf x}) f({\bf x})
= f({\bf x}).
\end{equation}
This definition leads to some interesting properties. First and foremost, we have the following simplifications:
\begin{equation}
\pi_i(u) = \frac{ f(T(u)) } {\sum_{j \in \mathcal{F}_k} p_j(T(u))} 
\label{eqn:SimplifiedWDistributions-1}
\end{equation}

Second, in each primary sample space the target distribution depends only on the path ${\bf x} = T(u)$, but not on the particular choice of technique $i$ used to generate it.
In other words, if $u^i \in U_i$ and $u^j \in U_j$ map to the same path ${\bf x} = T_i(u^i) = T_j(u^j)$, we have:
\begin{equation}
\pi_i(u^i) = \pi_j(u^j)
\end{equation}
In particular, the target distribution depends only on how well the \emph{sum}\footnote{Equivalently, their average, since $\pi$ is here defined up to a constant.} of the individual pdfs $p_i$ approximate $f$.
This is an interesting result, as we will see later on.

Third, notice that if all bidirectional techniques are included in $\mathcal{F}_k$, the target distribution does not contain any of the weak singularities induced by the geometric terms. This is the case because each pdf includes all but one of the geometric terms: thus their sum will contain all of them, and counterbalance those in the numerator of (\ref{eqn:SimplifiedWDistributions-1}). In particular, this means there will be no singular concentration of paths near geometric corners.\footnote{The only sources of singularie Diracs in unsampled specular BSDFs in SDS paths (not containing any DD edge).}
Notice that this would have not been the case if we simply adopted $\pi = f / p_i$, omitting the multiple importance sampling weight.

\vspace{2mm}
\subsection{Auxiliary Distributions}

\v{S}ik et al \shortcite{Sik:2016} proposed using an auxiliary distribution in conjunction with replica exchange \cite{Swendsen:1986} to help the primary MLT chain escape from local maxima. The auxiliary distribution is designed to be easier to sample, and hence favor exploration. Given they were working in the context of the original PSSMLT formulation where all connections are performed, they proposed using an auxiliary distribution with a target defined as 1 if any of the paths formed provides a non-zero contribution, and 0 otherwise.


With our new primary sample space formulation, a similar but even easier objective can be achieved by simply dropping all connection terms except for visibility, i.e. the only terms which are not sampled by the $i$-th local path sampling technique, giving:
\begin{equation}
\pi_i'(u) = V(x_{i-1} \leftrightarrow x_{i})
\end{equation}
which in path space becomes:
\begin{equation}
\bar{\pi_i}'(x) = V(x_{i-1} \leftrightarrow x_{i}) p_i(x)
\end{equation}
Notice that due to our use of primary sample space mutations, this function is very easy to sample, as our base sampling technique already generates samples distributed according to $p_i$.
Importantly, we might not even need Metropolis at all, as we could simply use our path generation technique as an independence sampler, akin to the \emph{large steps} in the original work of Kelemen et al \shortcite{Kelemen:2002}.
However, using Metropolis with local perturbations might still help in regions of difficult visibility.

\vspace{2mm}
\subsection{Handling color}

In the above we treated $f$ as a scalar, though in practice it is actually a color represented either in RGB or with some other spectral sampling.
While handling spectral rendering in all generality can require custom techniques \cite{Wilkie:2014:HWS} and is beyond the scope of this paper, for RGB (and even in many cases of spectral transport) it is sufficient to use the maximum of the components $f^* = \max_i\{(f)_i\}$ when constructing the target distribution, and weighting the resulting color samples accordingly before final image accumulation.

\vspace{10mm}

\section{Charted Metropolis-Hastings}

Before introducing our light transport algorithm, we introduce a novel family of general Markov chain Monte Carlo algorithms inspired by the primary sample space MLT formulation we just described. The idea is that we want to allow \emph{jumping} between different primary sample spaces, as this will allow to more freely escape from local maxima in situations in which the current parameterization is not the best fit for the target distribution.

Suppose in all generality that we have an arbitrary \emph{target space} $(\Omega,\mu)$, a function $f:\Omega \rightarrow \mathbb{R}$ we are interested in sampling, and a parametric family $\mathcal{F} = (U_i,T_i,R_i) _{i = 0,\dots,n-1}$,
such that:
\begin{description}
	\item $U_i$ is a measured \emph{primary sample space};
	\item $T_i$, the \emph{forward map}, is a function $T_i:U_i \rightarrow \Omega$;
	\item $R_i$, the \emph{reverse map}, is a right-inverse of $T_i$, i.e. $R_i:\Omega \rightarrow U_i$ with:
	\begin{equation}
	T_i(R_i(x)) = x \quad \forall x \in \Omega;
	\end{equation}
\end{description}


Let's also consider the density $p_i:\Omega \rightarrow \mathbb{R}$ defined as the pdf of the transformation $T_i(U)$ of a uniform random variable\footnote{More precisely, $p_i$ is uniquely defined almost everywhere as the function that satisfies the equation: $P(T_i(U) \in A) = \int_{A} p_i(x) d\mu(x)$, for any measurable subset $A \subseteq \Omega$ and $U \sim Uniform(U_i)$.}, and the function $r_i:U_i \rightarrow \mathbb{R}$ defined as its reciprocal:
\begin{equation}
r_i(u) = \frac{1}{p_i(T_i(u))}. \nonumber
\end{equation}
Now, consider again the weighted distributions defined by:
\begin{equation}
\pi_i(u) = \frac{ f(T_i(u)) }{ \sum_i p_i(T_i(u)) }
\label{eqn:WDistributions-2}
\end{equation}
The idea is that we could use the reverse maps $R_i$, which can be interpreted as inverse sampling functions, to perform the desired jumps between primary sample spaces, e.g performing swaps in the context of a replica exchange framework where we run $n$ chains, each sampled according to a different $\pi_i$. We now show how to achieve it.

\begin{figure}
  \fbox{\includegraphics[width=82.0mm]{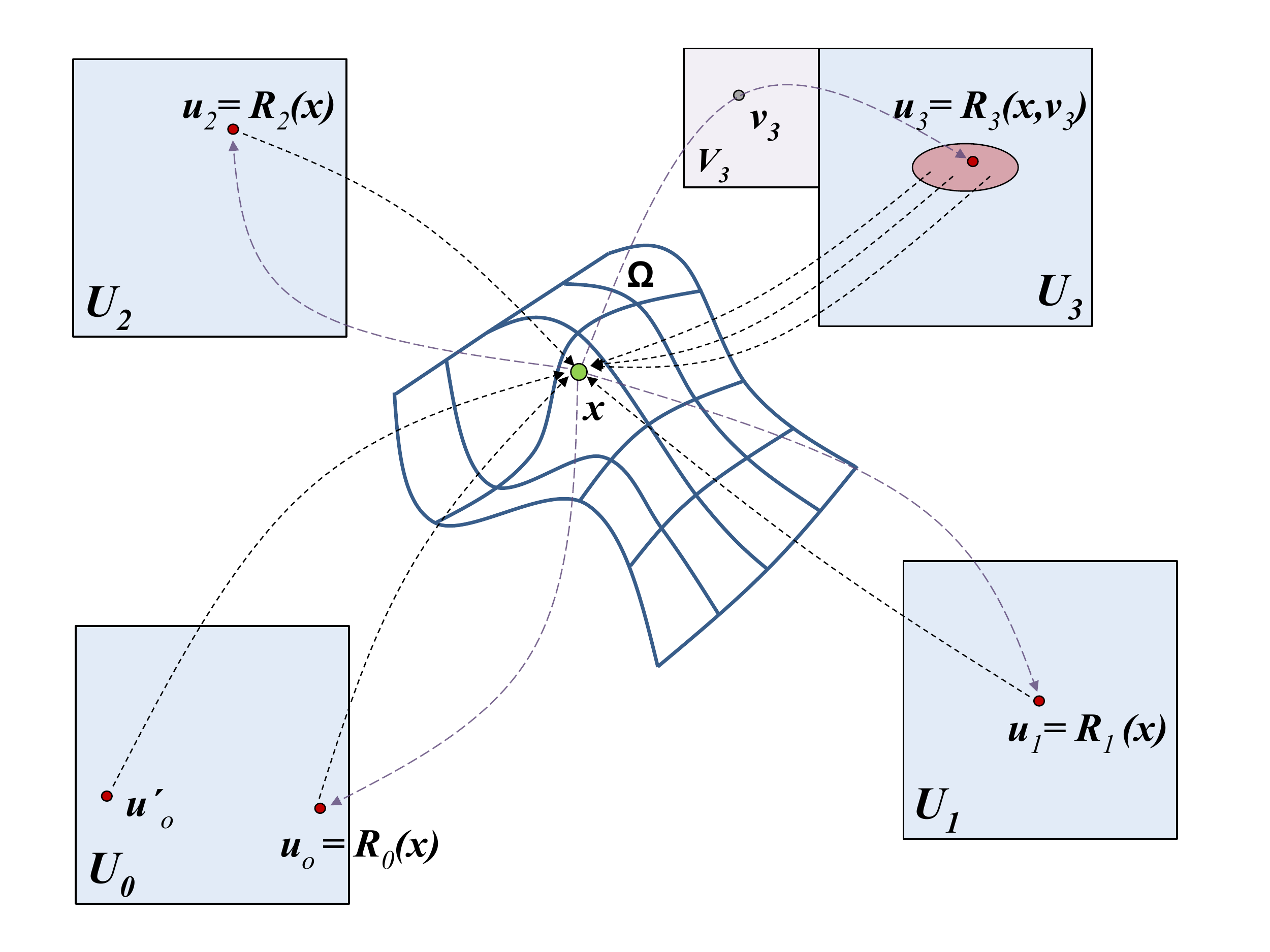}}
  \caption{Charted Metropolis-Hastings allows performing coordinate changes between the target space $\Omega$ and its sampling charts. When multiple points of a given sampling domain map to a single point in $\Omega$, it's sufficient for the right inversion mappings to return one of them (as for the case of $u_0$), or return one picked at random inside the set (as for the case of $u_3$) with the help of an additional sampling domain ($V_3$, light violet box). }
  \label{CMH}
\end{figure}

Given two states, $u_1^i$, generated by the $i$-chain, and $u_2^j$, generated by the $j$-chain, consider their target space mappings:
\begin{eqnarray}
{x_1} := T_i(u_1^i) \nonumber \\
{x_2} := T_j(u_2^j) \nonumber
\end{eqnarray}
and their \emph{reverse} mappings:
\begin{eqnarray}
{u_1^j} := R_j(x_1) \nonumber \\
{u_2^i} := R_i(x_2) \nonumber
\end{eqnarray}
if we wanted to perform a swap, preserving detailed balance between the chains requires accepting the swap with probability:
\begin{equation}
A =
\min \left( 1,
\frac{ 
	\pi_i(u_2^i)
	\pi_j(u_1^j)
	r_i(u_1^i)
	r_j(u_2^j)
} {
	\pi_i(u_1^i)
	\pi_j(u_2^j)
	r_i(u_2^i)
	r_j(u_1^j)
}
\right)
\label{eqn:PTAcceptanceRatio}
\end{equation}
This can be proven by looking at the two chains as an ensemble in the space $U_i \times U_j$, with target distribution $\pi_i \cdot \pi_j$. Equation (\ref{eqn:PTAcceptanceRatio}) is then obtained from equation (8) following the usual Metropolis-Hastings rule described in section 3.1, viewing $(u_1^i,u_2^j)$ as the current state and $(u_2^i,u_1^j)$ as the proposal.

In the previous section we saw that our target distributions $\pi_i$ assume the same value on the same points of $\Omega$, independently of the underlying technique $i$ used to generate it. Now since $R_i$ has been defined as a right inverse of $T_i$, if $u^j = R_j(T_i(u^i))$, we would again have:
\begin{equation}
\pi_j(u^j) = \pi_i(u^i).
\end{equation}
This property is essentially stating that our target distribution is invariant under a change of charts of the target space.

Hence, equation (\ref{eqn:PTAcceptanceRatio}) simplifies to:
\begin{equation}
A =
\min \left( 1,
\frac{ 
	r_i(u_1^i)
	r_j(u_2^j)
} {
	r_i(u_2^i)
	r_j(u_1^j)
}
\right)
\end{equation}
without requiring any evaluation of the target distributions.
Notice that we didn't require the transformations $T_i$ to be fully invertible: if the fiber of $x$ under $T_i$, i.e. the set $T^\leftarrow_i(x) = \{u | T_i(u) = x\}$, contains several points, it's sufficient that $R_i$ returns one of them.
This approach is very general, as such a function can always be constructed.
However, it can be made even more general by \emph{randomizing} the selection of the point in the fiber.
We do so by extending the domains in which the functions $R_i$ operate.
\paragraph{{\bf Definition}:} Sampling Atlas.
We call \emph{sampling atlas} a family $\mathcal{F} = (U_i,V_i,T_i,R_i)_{i=0,...,n-1}$ where $U_i$ and $T_i$ are defined as before, but:
\begin{description}
	\item $V_i$ is a measured \emph{reverse sampling space}, and
	\item $R_i$ is an \emph{extended right-inversion map}, $R_i:\Omega \times V_i \rightarrow U_i$, such that:
	\begin{equation}
	T_i(R_i(x,v)) = x \quad \forall x \in \Omega \quad \textrm{and} \quad \forall v \in V_i. \nonumber
	\end{equation}
\end{description}
Each tuple $(U_i,V_i,T_i,R_i)$ is called a \emph{sampling chart}.

\vspace{2mm}
With these definitions, we can draw two uniform random variables $v_1 \in V_i$ and $v_2 \in V_j$, and replace the reverse mappings $u_1^j$ and $u_2^i$ with:
\begin{eqnarray}
{u_1^j} := R_j(x_1,v_1) \nonumber \\
{u_2^i} := R_i(x_2,v_2) \nonumber
\end{eqnarray}
\ifnum 1 = 0
which can now be tested for acceptance with the same acceptance ratio $A$ as defined in equation 22.
\else
which can now be tested for acceptance with the same acceptance ratio:
\begin{equation}
A =
\min \left( 1,
\frac{ 
	r_i(u_1^i)
	r_j(u_2^j)
} {
	r_i(u_2^i)
	r_j(u_1^j)
}
\right). \nonumber
\end{equation}
\fi

This construction is depicted in Figure~\ref{CMH}, where:
a. the chart $U_0$ contains two points, $u_0$ and $u'_0$, that map to the same point ${\bf x} \in \Omega$, but $R_{0}({\bf x})$ selects just one of them, in this case $u_0$;
b. the chart $U_3$ contains an entire set that maps to $x$, but its points are identified by means of points of the reverse sampling domain $V_3$.
	
\vspace{2mm}
A similar mathematical framework can be used in the context of serial (or simulated) tempering \cite{Marinari:1992}. In this context, one could run a single chain $u^i = (u,i)$ in an extended state space $U \times \mathcal{F}$, where $i$ denotes the technique used to map the chain to target space.
Drawing a uniform random variable $v \in V_i$ and swapping from $i$ to $j$ through the transformation:
\begin{equation}
u^j = R_j(u^i,v) \nonumber
\end{equation}
would then require accepting the swap with probability:
\begin{equation}
\min \left( 1,
\frac{ 
	r_i(u^i)
} {
	r_j(u^j)
}
\right)
\label{eqn:STAcceptanceRatio}
\end{equation}
and rejecting it otherwise.
Once again, no evaluation of the target distributions is required.
We call both this and the above mutations \emph{chart swaps} or \emph{coordinate changes}.

Notice that if there is a way to craft mutations in the target space itself, it is always possible to add the identity chart to $\mathcal{F}$:
\begin{description}
	\item $U_n = \Omega$, $V_n = \emptyset $
	\item $T_n(x) = R_n(x) = x$;
\end{description}
care must only be taken in adding the probability $p_n = 1$ to the denominator of all the distributions $\pi_i$ in equation (\ref{eqn:WDistributions-2}).

Finally, we consider another type of mutation, \emph{inverse primary space perturbations}, which can be in a sense considered the dual of the above. Suppose we are now running a chain in the target space $\Omega$, distributed according to $\pi({\bf x})$.
We can then use inversion to momentarily parameterize the target space through a given technique $i$ and take a detour or \emph{move down} from $\Omega$ to $U_i$ to perform a symmetric primary sample space perturbation there, before finally getting back to $\Omega$.
With this scheme, given a state ${\bf x}$ and a uniform random variable $v \in V_i$, applying the transformation $R_i$ to obtain $u = R_i({\bf x},v)$ and the perturbation kernel $K$ to obtain the proposal $u' = K(u)$ and ${\bf y} = T_i(u')$, would result in the following acceptance ratio:
\begin{equation}
A({\bf y}|{\bf x}) =
\min \left( 1,
\frac{
	\pi({\bf y})K(u|u')r_i(u')
}
{
	\pi({\bf x})K(u'|u)r_i(u)
}
\right)
\end{equation}
which simplifies to the standard primary sample space formula if $K$ is symmetric:
\begin{equation}
A({\bf y}|{\bf x}) =
\min \left( 1,
\frac{
	\pi({\bf y})
}
{
	p_i({\bf y})
}
\cdot
\frac{
	p_i({\bf x})
}
{
	\pi({\bf x})
}
\right).
\end{equation}

We call this family of MCMC algorithms that jump between charts of the target space \emph{charted Metropolis-Hastings}, or CMH.

\begin{figure*}
	\fbox{\includegraphics[width=170.0mm]{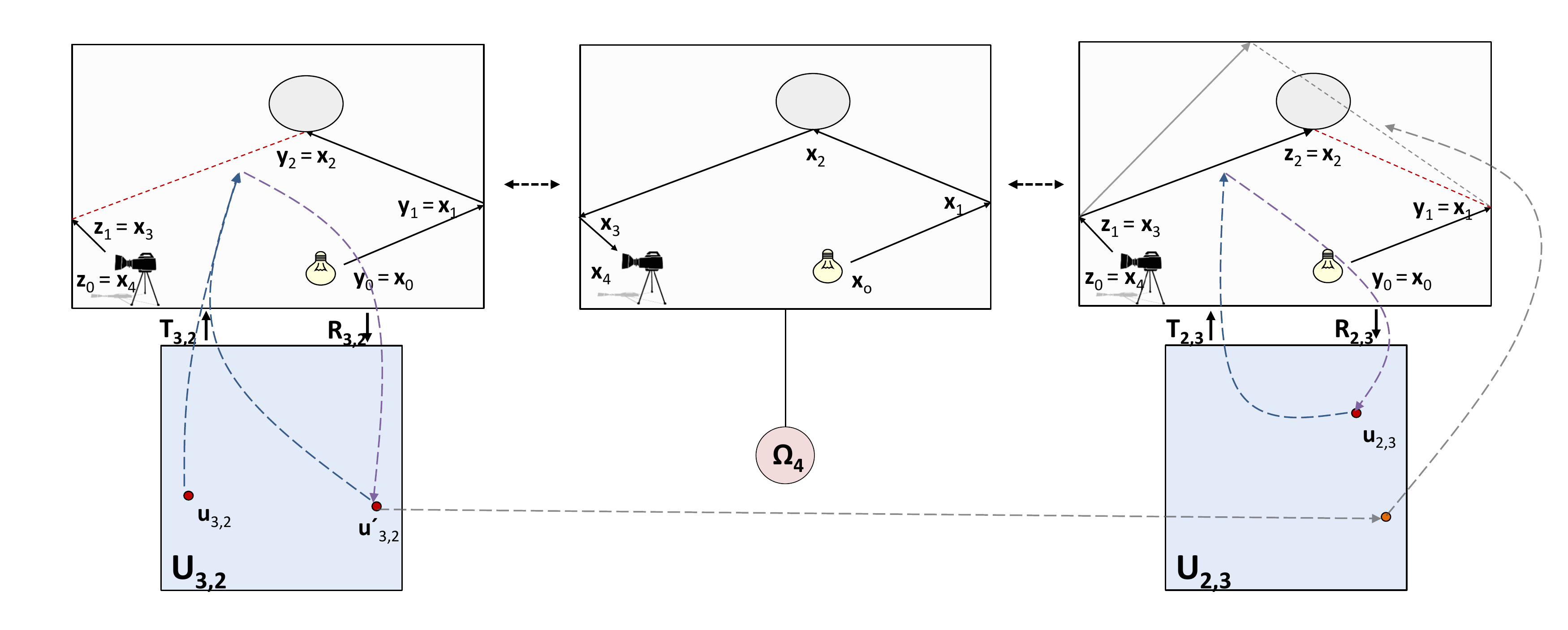}}
	\caption{A visualization of two path space charts,
	where one of the bidirectional sampling techniques, in this case $T_{3,2}$, maps multiple points to the same selected path, while $T_{2,3}$ is locally invertible.
	 Notice how a naive transfer of coordinates such as that employed in MMLT (dashed gray lines) could result in a very different path.
	}
	\label{CMLT-fig}
\end{figure*}

\section{Charted Metropolis Light Transport}

It should now be clear how the above algorithms can be applied to light transport simulation.
If we consider the framework for primary sample space MLT outlined in section 3.2, it is sufficient to add functions for \emph{path sampling inversion} to be able to apply our new charted Metropolis-Hastings replica exchange or serial tempering mutations in conjunction with the standard set of primary sample space perturbations.
The advantage of these mutations is that they will allow to more easily escape from local maxima when the current sampling technique is not locally the best fit for $f$.
The mutations are relatively cheap, as they don't require any expensive evaluations of the target distribution.

Moreover, and very importantly, the algorithm is made practical by not requiring the path sampling functions $T_i$ to be classically invertible.
In the context of light transport simulation this property is crucial, as BSDF sampling is seldom invertible: in fact, with layered materials often a random decision is taken to decide which layer to sample, but the resulting output directions could be equally sampled (with different probabilities) by more than one layer.
Our framework requires to return just one of them, but it also allows selecting which one at random with a proper probability. All is needed is the ability to compute the density of the resulting transformation.
This construction is illustrated in Figure~\ref{CMLT-fig}, which shows how the same path ${\bf x}$ can be represented both in the chart corresponding to the bidirectinal technique $(2,3)$ and the one coresponding to the technique $(3,2)$, where the latter contains two distinct points, $u_{3,2}$ and $u'_{3,2}$, that map to ${\bf x}$. In the picture $R_{3,2}({\bf x})$ selects just one of them, in this case $u'_{3,2}$. 

Further on, by adding the identity target space chart, we can also add the original path space mutations proposed by Veach and Guibas \shortcite{Veach:1997:MLT}, potentially coupled with the new inverse primary space perturbations.

We call the family of such algorithms \emph{charted Metropolis light transport}, or CMLT.

\subsection{Connection to path space MLT}


The new algorithms can be considered as a bridge between primary sample space MLT and the original path space MLT proposed by Veach and Guibas \shortcite{Veach:1997:MLT}.
In fact, one of the advantages of the original formulation over Kelemen's variant \shortcite{Kelemen:2002} was its ability to \emph{break the path in the middle} and resample the given path segment with any arbitrary bidirectional technique.
This ability was entirely lost in primary sample space, as the bidirectional sampling technique was implicitly determined by the sample coordinates (or needed to be chosen ahead of time in the version we outlined in section 3.2).
While Multiplexed Metropolis Light Transport (MMLT) \cite{Hachisuka:2014} added the ability to change technique over time, as the coordinates $u$ were kept fixed such a scheme was leading to swap proposals that sample unrelated paths: in fact, two techniques $i$ and $j$ map the same coordinates $u$ to different paths $T_i(u) \neq T_j(u)$ that share only a portion of their prefixes (in other words, the two resulting paths are \emph{spuriously} correlated by the algorithm, whereas in fact there is no reason for them to be - see Fig.\ref{CMLT-fig}).
Our coordinate changes, in contrast, \emph{preserve} the path while changing its parameterization, thus allowing to simply perturb it later on with a different bidirectional sampler.

Adding the identity path space chart and inverse primary space perturbations makes the connection even tighter, allowing to smoothly integrate the original bidirectional mutations and perturbations with an entirely new set of primary sample space perturbations.

Notice that while inverse primary space perturbations could also be applied to a single path space chain, the advantage of also incorporating primary space chains in a replica exchange or serial tempering context is that the target distributions (defined by equation~\ref{eqn:WDistributions-2}) become generally smoother due to the implicit use of the multiple importance sampling weight, raising the acceptance rate.

\subsection{Alternative parameterizations}

While the original primary sample space Metropolis used path space parameterizations based on plain BSDF sampling, it is also possible to use other parameterizations that can provide further advantages: for example the half vector space parameterizations that have been recently explored \cite{Kaplanyan:2014:HSLT,Hanika:2015:IHLST}.

\subsection{Density estimation}

So far, we have concentrated on standard bidirectional path tracing with vertex connections. However, all the above extends naturally to density estimation methods, using the framework outlined in \cite{Hachisuka:2012}.
The only major difference is the computation of the subpath probabilities.


However, we here suggest an alternative approach.
Instead of using density estimation as an additional technique, applying multiple importance sampling to combine it into a unique estimator, we can use it only to craft additional proposals.
In other words, we can use density estimation as another independence sampler.
Suppose we are running an MCMC simulation in $\Omega_k$, and at some point in time our chain is in the state $u^i$, with $s = i$, and $t = k - s + 1$.
We can then try to build a candidate path through density estimation with the $(s+1,t)$-technique and, if the resulting path has non-zero contribution, we can drop one light vertex (and the corresponding primary sample space coordinates) and consider it as a new proposal $u^i_{de}$.
Notice that in doing so, we have to adjust the acceptance ratio for the actual  proposal distribution. For clarity, we will now omit the superscripts $i$, and obtain:
\begin{equation}
A(u_{de}|u) = \min \left( 1,
\frac{
	\pi(u_{de})p_{de}(T(u))
}
{
	\pi(u)p_{de}(T(u_{de}))
} \right)
\end{equation}
where $p_{de}(x)$ is the probability of sampling the path $x$ by density estimation (which can be approximated at the cost of some bias as described in \cite{Hachisuka:2012} or estimated unbiasedly as in \cite{Qin:2015:UPG}).

If we want to further raise the acceptance rate, we can also mix this proposal scheme with an independence sampler based on bidirectional connections and combine the two, calculating the total expected probability to make both more robust:
\begin{equation}
A(u'|u) = \min \left( 1,
\frac{
	\pi(u')(p_{de}(T(u)) + p_{bc}(T(u)))
}
{
	\pi(u)(p_{de}(T(u')) + p_{bc}(T(u')))
}
\right).
\end{equation}
Notice that this formula is now agnostic of how the samples were generated in the first place, i.e. whether the candidate $u'$ was proposed by density estimation or bidirectional connections: this is a positive side-effect of using expectations.\footnote{While this looks similar to multiple importance sampling, it is not quite the same: multiple importance sampling is a more general technique used to combine estimators, whereas here we are just interested in computing an expected probability density, using so called state-independent mixing \cite{Geyer:2011}. However, multiple importance sampling using the balance heuristic is equivalent to using an estimator based on the average of the probabilities, which is exactly the expected probability we need: hence the reason of the similarity. This approach is the same used in the original MLT to compute the expected probability of bidirectional mutations.}




\begin{figure}
	\fbox{\includegraphics[width=82.0mm]{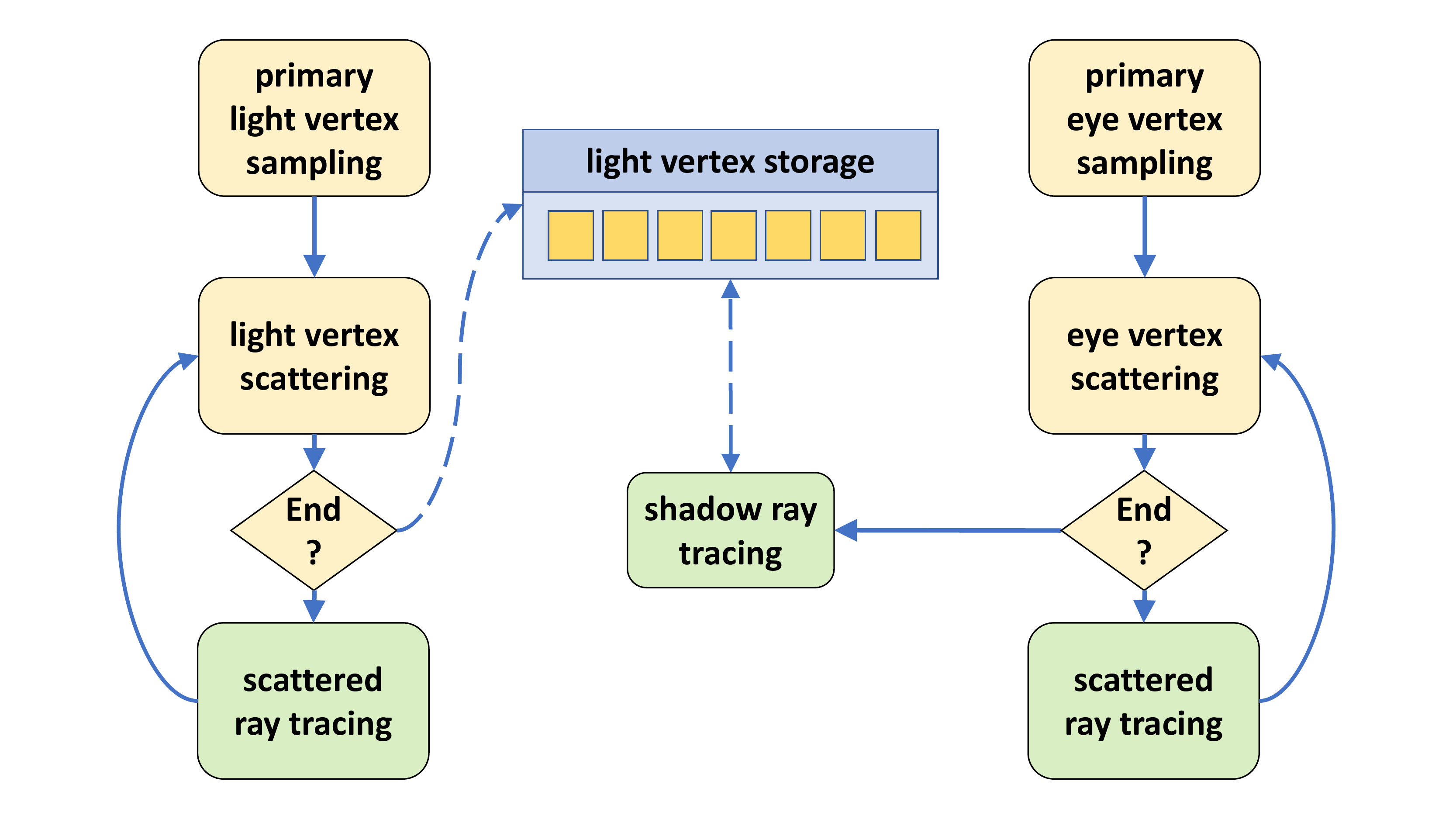}}
	\caption{A schematic visualization of the basic bidirectional path tracing pipeline, showing the different shading and tracing kernels. Notice that while they are shown here side by side, light path tracing and eye path tracing happen in subsequent phases of the algorithm. }
	\label{PT}
\end{figure}

\vspace{8mm}

\subsection{Designing a complete algorithm}

So far we have only constructed a theoretical background to build novel algorithms, but we didn't prescribe practical recipes.
The way we combine all the above techniques into an actual algorithm is described here.

First of all, we start by estimating the total image brightness with a simplified version of bidirectional path tracing.
The algorithm first traces $N_{init}$ light subpaths in parallel and stores all generated light vertices.
Then it proceeds tracing $N_{init}$ eye subpaths, and connects each eye vertex to a single light vertex chosen at random among the ones we previously stored.
At the same time, the emission distribution function at each eye vertex is considered, forming pure path tracing estimators with light subpaths with zero vertices.
All evaluated connections (both implicit and explicit) with non-zero contribution (which represent entire paths, each with a different number of light and eye vertices $s$ and $t$) are stored in an unordered list.

Second, in order to remove startup bias, we resample a population of $N$ seed paths for a corresponding amount of chains.
In order to do this, we build the cumulative distribution over the scalar contributions of the previously stored paths, and resample $N$ of them randomly.

Notice that the $N$ seed paths will be distributed according to their contribution to the image. Particularly, the number of paths sampled with technique $i$ will be proportional to the overall contribution of that technique, and similarly for path length.
At this point, though not crucial for the algorithm, we sort the seeds by path length $k$ so as to improve execution coherence in the next stages.
In practice, sorting divides the $N$ seeds into groups of $N_k$ paths each, such that $\sum_k N_k = N$.

Finally, we run the $N$ Markov chains in parallel using both classic primary sample space perturbations and the novel simulated tempering or replica exchange mutations described in sections 3 and 4. As the new mutations have a low cost compared to performing actual perturbations, they can be mixed in rather frequently (with very low overhead up to once every four iterations).

\RestyleAlgo{boxruled}
\begin{algorithm}
	\KwData{x, $\omega_i$, $\omega_o$}
	\KwResult{u (primary space coordinates)}
	{
		probs[] $\leftarrow$
		layerSamplingProbabilities(x,$\omega_i$)\;
		
		prob\_sum $\leftarrow$ probs[diffuse] + probs[glossy]\;
		
		v $\leftarrow$ random() * prob\_sum\;
		
		\eIf{v $<$ probs[diffuse]}{
			u $\leftarrow$ (v, invertLambert(x,$\omega_i$,$\omega_o$))\;
		}{
			u $\leftarrow$ (v, invertGGX(x,$\omega_i$,$\omega_o$))\;
		}
	}
	\caption{inversion of a composite BSDF containing a diffuse and a glossy layer}
\end{algorithm}

\section{Implementation}

We implemented our algorithm, together with MMLT, PSSMLT and bidirectional path tracing (BPT) in CUDA C++, exposing massive parallelism at every single stage, including ray tracing, shading, cdf construction (prefix sum), resampling and sorting (radix sort).
The basic bidirectional path tracing algorithm is constructed as a pipeline of kernels (also known as wavefront tracing \cite{Laine:2013:MCH}), and relies on the OptiX Prime library for ray tracing.
We ran all tests on an NVIDIA Maxwell Titan X GPU.

The basic bidirectional path tracing pipeline, composed of seven shading and tracing stages, is shown schematically in Figure~\ref{PT}.
This pipeline is further extended in all the MCMC rendering algorithms by additional stages performing primary sample space coordinates generation (applying both perturbations and chart swaps in the case of CMLT), and the final acceptance-rejection step.
All the pipeline stages communicate through global memory work queues.

In order to keep storage and bandwidth consumption to a minimum, only minimal information is stored for each path vertex (including instance id, primitive id and uv coordinates), using on the fly vertex attributes interpolation where needed (such as during path inversion).
For $256K$  paths, of a maximum of 10 vertices each, this requires about 64MB of storage.

Both our CMLT and MMLT implementations run several thousand chains in parallel, using the seeding algorithm described in section 5.4.
Besides being strictly necessary to scale to massively parallel hardware, we found this to produce some additional image stratification, as discussed in the Results section.
The CMLT implementation is based on the serial tempering formulation.

Our framework employs a layered material system that combines a diffuse BSDF (Lambertian) and rough glossy reflection and transmission BSDFs (GGX) using a Fresnel weighting.
Sampling of the glossy component is implemented using the distribution of visible normals \cite{heitz:hal-2014}, and selection between the diffuse and glossy components is performed based on Fresnel weights.
Clearly, this path sampling scheme is not invertible, as both the diffuse and glossy components can map different primary sample space values to the same outgoing directions.
Hence, we used the machinery described in section 4 to enable randomized inversion.


\subsection{Chart swaps and path inversion}

Given a bidirectional path generated by the technique $(s,t)$ using coordinates $u$, in order to perform a chart swap we propose a new pair $(s',t')$ with $s' + t' = s + t$ distributed according to the total energy of the techniques (i.e. the normalization constants of the target distributions).
After the candidate is sampled, path inversion needs to be performed using the transformation $u' = R_{s',t'}(T_{s,t}(u))$.
This transformation can be widely optimized noticing that there are only two cases:
\begin{description}
	\item $s' > s$: in this case it is only necessary to invert the coordinates of the light subpath vertices $\{y_s, ..., y_{s' - 1}\}$.
	\item $t' > t$: in this case it is only necessary to invert the coordinates of the eye subpath vertices $\{z_t, ..., z_{t' - 1}\}$.
\end{description}
Computing the inverse pdf $r_{s,t}$ can be optimized analogously.

In each of these cases, we start the stochastic inversion from the end of the selected subpath, and proceed backwards. At each vertex, we consider the local composite BSDF, and compute the forward probabilities originally used to select which layer to sample (for example, based on their Fresnel weighted albedos). Using these, we draw a single random number $v$ to select which of the layers to use for inversion. Pseudocode for a material with a diffuse and glossy layer is provided in Algorithm 1.
Pseudocode for a serial version of the overall CMLT algorithm is given in Algorithm 2.
The Appendix provides further details and pseudocode for the inversion of typical BSDFs.

\section{Results}

\ifnum 1 = 0

\begin{figure*}
	\centering
	\includegraphics[width=42.0mm]{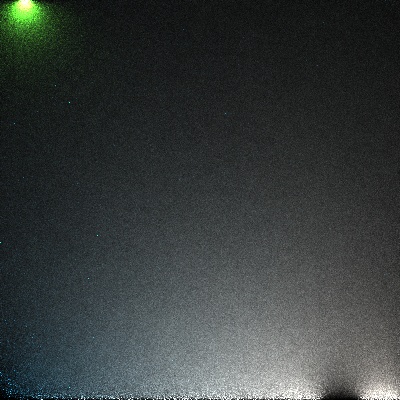}
	\includegraphics[width=42.0mm]{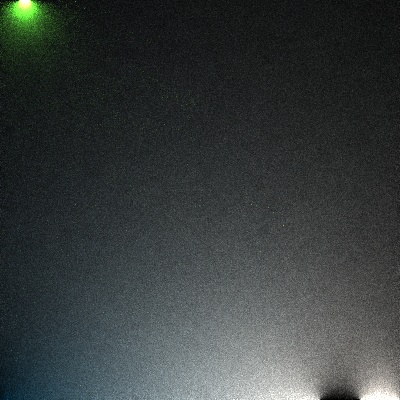}
	\includegraphics[width=42.0mm]{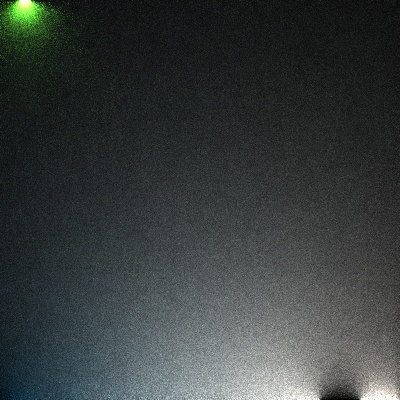}
	\includegraphics[width=42.0mm]{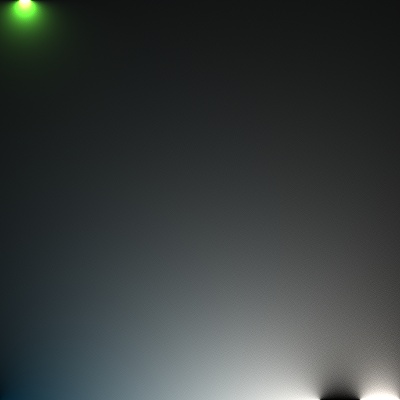} \\
	\includegraphics[width=42.0mm]{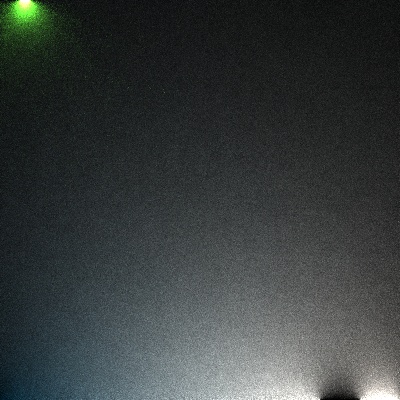}
	\includegraphics[width=42.0mm]{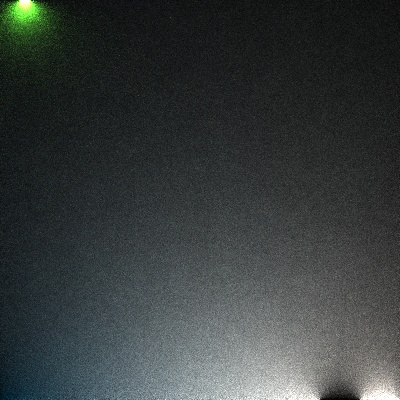}
	\includegraphics[width=42.0mm]{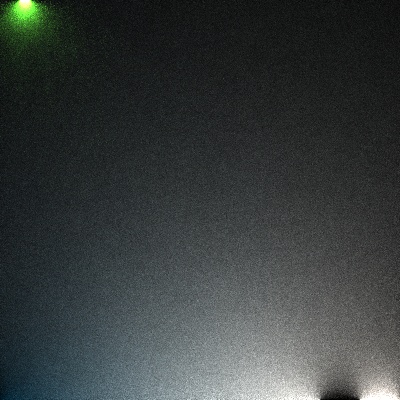}
	\includegraphics[width=42.0mm]{results/test_ref}
	\caption{A simple test scene featuring direct lighting from two equally sized area lights: the first, with spatially varying emission properties, partially blocked by a thin black strip; the second only reachable through a tiny hole. From left to right, first row:  PSSMLT-1, PSSMLT-2, PSSMLT-AVG, reference image. Second row: PSSMLT-MIX, CMLT-IPSM, CMLT, reference image. }
	\label{CMH-tests}
\end{figure*}

\else

\begin{figure}
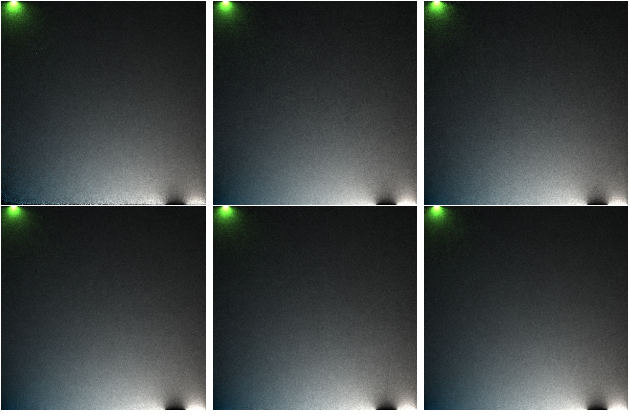

	\centering
	\includegraphics[width=27.0mm]{results/test_kmlt1_low}
	\includegraphics[width=27.0mm]{results/test_kmlt2_low}
	\includegraphics[width=27.0mm]{results/test_kmlt_avg_low} \\
	\includegraphics[width=27.0mm]{results/test_kmlt_mix_low}
	\includegraphics[width=27.0mm]{results/test_cmlt_epsm_low}
	\includegraphics[width=27.0mm]{results/test_cmlt_low}
	\caption{
		A simplified light transport test.
		From left to right, first row:  PSSMLT-1, PSSMLT-2, PSSMLT-AVG. Second row: PSSMLT-MIX, CMLT-IPSM, CMLT. }
	\label{CMH-tests}
\end{figure}

\fi

We performed two sets of tests.
The first is aimed at testing the many possible algorithmic variations of CMLT on a simplified light transport problem.
The second, using full light transport simulation, compares a single CMLT variant against MMLT, which could be currently considered state-of-the-art in primary sample space MLT.

\subsection{Simplified light transport tests}

This test consists of rendering an orthographic projection of the $XY$ plane directly lit by two area light sources.
The first light is a unit square on the plane $Y = 0$, with a spatially varying emission distribution function changing color and increasing in intensity along the $X$ axis. The light source is partially blocked by a thin black vertical strip near its area of strongest emission.
The second light is another unit square on the plane $Y = 1$, with uniform green emission properties. This light is completely blocked except for a tiny hole.

In this case, our path space consists of two three-dimensional points: the first on the ground plane, the second on the light source.
As charts, we used two different parameterizations:
\paragraph{{\bf 1.}}
generating a point uniformly on the visible portion of the ground plane and a point on the light sources distributed according to their spatial emission kernels - corresponding to path tracing with next-event estimation, i.e. the bidirectional path tracing technique $(s,t) = (1,1)$;
\paragraph{{\bf 2.}}
generating a point uniformly on the visible portion of the ground plane, sampling a cosine distributed direction, and intersecting the resulting ray with the scene geometry to obtain the second point - corresponding to pure path tracing, i.e. the  bidirectional path tracing technique $(s,t) = (0,2)$.

\vspace{2mm}
Both charts have a four dimensional domain, and in both cases we used exact inverses of the sampling functions.
We tested six different MCMC algorithms:

\begin{description}
	\item {\bf PSSMLT-1}: a single PSSMLT chain using the first parameterization;
	\item {\bf PSSMLT-2}: a single PSSMLT chain using the second parameterization;
	\item {\bf PSSMLT-AVG}: two PSSMLT chains using both the first and the second parameterization, both using the importance sampled distribitions (equation~\ref{eqn:ISDistributions}), where the accumulated  image samples are weighted (i.e. averaged) through multiple importance sampling with the balance heuristic;
	\item {\bf PSSMLT-MIX}: two PSSMLT chains using both the first and the second parameterization, with the weighted distributions (equation~\ref{eqn:WDistributions-1});
	\item {\bf CMLT-IPSM}: a single CMLT chain in path space alternating inverse primary space mutations using the first and the second parameterizations;
	\item {\bf CMLT}: two CMLT chains using both the first and the second parameterization as charts, coupled with replica-exchange swaps performed every four iterations;
\end{description}

Results are shown in Figure~\ref{CMH-tests}, while their root mean square error (RMSE) is reported in Table 1.
All images except for the reference were produced using the same total number of samples $n = 16 \cdot 10^6$: PSSMLT-1, PSSMLT-2 and CMLT-IPSM running a single chain of length $n$, whereas PSSMLT-AVG, PSSMLT-MIX and CMLT running two chains of length $n/2$.
In table 1 we further report RMSE values for $n = 128 \cdot 10^6$.
The reference image has been generated by plain Monte Carlo sampling.

As can be noticed, our PSSMLT-MIX formulation using the distributions defined by equation (\ref{eqn:WDistributions-1}) is superior to simply averaging two PSSMLT chains using multiple importance sampling (PSSMLT-AVG), which is in fact worse than PSSMLT using a single chain according to the second distribution (PSSMLT-2).

\begin{table}[!t]
	\begin{center}
		\footnotesize
		\begin{tabular}{|l|c|c|}
			\hline
			Algorithm & n = 16 $\cdot$ 10e6 & n = 128 $\cdot$ 10e6\\
			\hline
			\texttt{PSSMLT-1}   	&  8.287 $\cdot$ 10e{-2} & 3.212 $\cdot$ 10e{-2} \\\hline
			\texttt{PSSMLT-2} 		&  4.073 $\cdot$ 10e{-2} & 1.488 $\cdot$ 10e{-2} \\\hline
			\texttt{PSSMLT-AVG}   	&  4.106 $\cdot$ 10e{-2} & 1.587 $\cdot$ 10e{-2} \\\hline
			\texttt{PSSMLT-MIX} 	&  3.663 $\cdot$ 10e{-2} & 1.405 $\cdot$ 10e{-2} \\\hline
			\texttt{CMLT-IPSM}      &  3.924 $\cdot$ 10e{-2} & 1.467 $\cdot$ 10e{-2} \\\hline
			\texttt{CMLT} 			&  {\bf 3.502 $\cdot$ 10e{-2}} &  {\bf 1.374 $\cdot$ 10e{-2}}\\\hline
		\end{tabular}
	\end{center}
	\caption{Root mean square error of the images computed by the various algorithms we tested in figure~\ref{CMH-tests}.}
	\label{tab:breakdown}
\end{table}

CMLT-IPSM produces results that are just slightly worse than PSSMLT-MIX, but still superior to all other PSSMLT variants.
The reason why CMLT-IPSM is inferior to PSSMLT-MIX is that while the target distribution for CMLT-IPSM is proportional to $f$, the target distributions of the chains in PSSMLT-MIX are smoother due to the embedded multiple importance sampling weights, and contain no singularities.

Finally, CMLT produces the best results among all algorithms.

\subsection{Full light transport tests}

For these tests we compared the CMLT implementation described in section 5 against our own implementation of MMLT.
We provide five test scenes representative of different transport phenomena:

\begin{description}
	\item {\bf The Gray \& White Room}: a scene from Bitterli's repository \shortcite{resources16}.
	
	\item {\bf Escher's Room}: an M.C. Escher themed adaptation of the above scene, featuring multi-layer materials with variable surface properties. This scene contains many light sources of different size: the large back wall, with a variable Lambertian emission distribution displaying a famous painting by the artist, a smaller area light on the ceiling, and the external lighting coming from the windows. The smaller light is partially blocked by a rough glossy reflector, which causes a blurry caustic on the partially glossy ceiling.
	Notice how all the above elements contribute to forming an \emph{all-frequency} lighting scenario.
	
	\item {\bf Escher's Glossy Room}: a variation of the above scene in which all surfaces are glossy (with no diffuse component), with variable roughness (with GGX exponents ranging between 5 and 100). Notice that this scene contains a variety of caustics of all frequencies (in a sense, all lighting is due to caustics).
	This scene stresses the advantages of chart swaps in the presence of near-specular transport, where there are many narrow modes and there is often no single best sampling technique.
	
	\item {\bf Wall Ajar}: another variation of the above scene mimicking Eric Veach's famous scene \emph{the door ajar}. Most of the lighting in the scene comes from a narrow opening in the \emph{sliding} back wall, which covers an equally large but completely hidden emissive wall. Hence, the room is almost entirely indirectly lit, except for the blue light coming from the windows. The ceiling area light source is also considerably smaller, casting a sharper caustic, and most surfaces are now about half diffuse half glossy.
	The sofa also features some rough transmission.

	\item {\bf Salle de bain}: another scene from Bitterli's repository \shortcite{resources16}. While in terms of light transport this scene is considerably simpler than any of the others, we chose it as representative of some typical architectural lighting situations.
\end{description}

It is important to note that while the first four scenes look superficially similar, they stress entirely different transport phenomena. Moreover, all of them, while relatively simple in terms of geometric complexity, are very hard in terms of pure light transport, requiring between $16 \cdot 10^3$ and $64 \cdot 10^3$ samples per pixel (spp) for bidirectional path tracing to converge.

\begin{figure*}
	\centering
	
\begin{subfigure}[b]{0.325\textwidth}
	\centering
	\includegraphics[width=58.0mm]{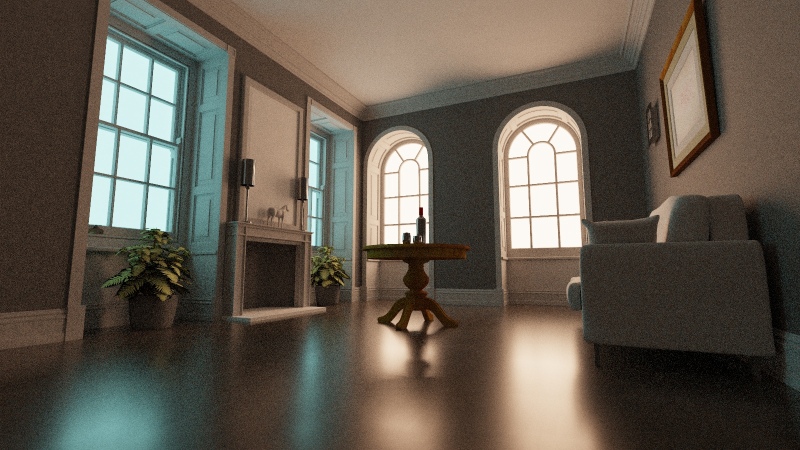}
\end{subfigure}
\begin{subfigure}[b]{0.325\textwidth}
	\centering
	\includegraphics[width=58.0mm]{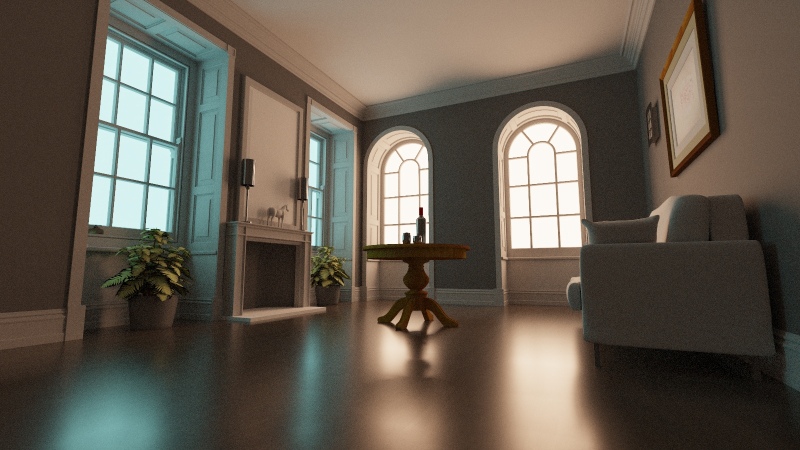}
\end{subfigure}
\begin{subfigure}[b]{0.325\textwidth}
	\centering
	\includegraphics[width=58.0mm]{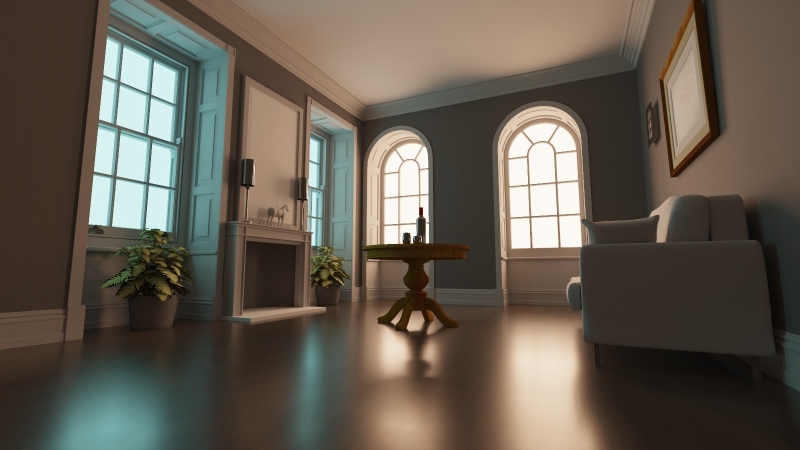}
\end{subfigure}
\\
%
%
\begin{subfigure}[b]{0.1595\textwidth}
	\centering
	\includegraphics[width=28.5mm]{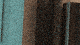}
\end{subfigure}
\begin{subfigure}[b]{0.1595\textwidth}
	\centering
	\includegraphics[width=28.5mm]{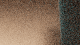}
\end{subfigure}
\begin{subfigure}[b]{0.1595\textwidth}
	\centering
	\includegraphics[width=28.5mm]{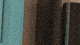}
\end{subfigure}
\begin{subfigure}[b]{0.1595\textwidth}
	\centering
	\includegraphics[width=28.5mm]{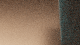}
\end{subfigure}
\begin{subfigure}[b]{0.1595\textwidth}
	\centering
	\includegraphics[width=28.5mm]{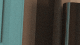}
\end{subfigure}
\begin{subfigure}[b]{0.1595\textwidth}
	\centering
	\includegraphics[width=28.5mm]{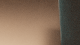}
\end{subfigure}
\\
\begin{subfigure}[b]{0.325\textwidth}
	\centering
	\includegraphics[width=58.0mm]{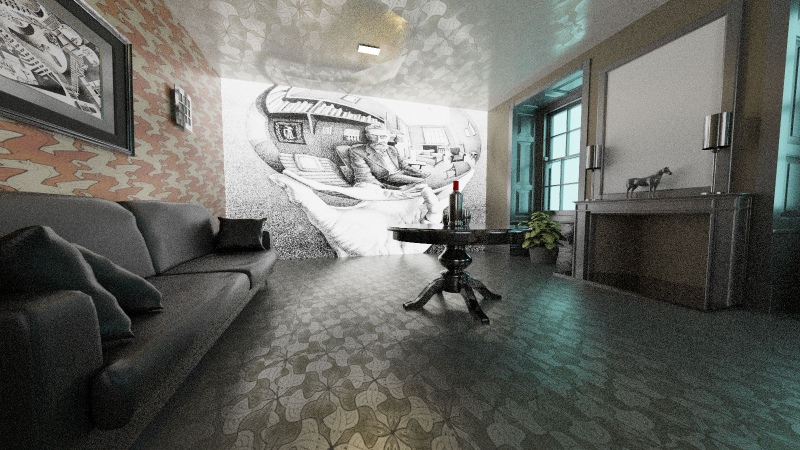}
\end{subfigure}
\begin{subfigure}[b]{0.325\textwidth}
	\centering
	\includegraphics[width=58.0mm]{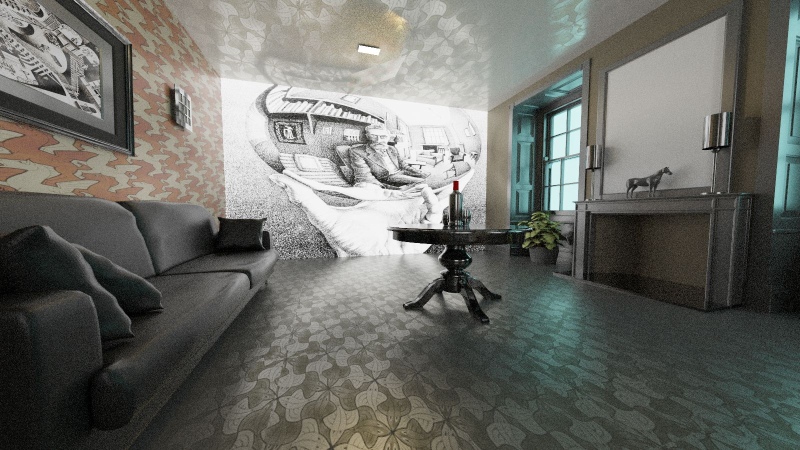}
\end{subfigure}
\begin{subfigure}[b]{0.325\textwidth}
	\centering
	\includegraphics[width=58.0mm]{results/escher-room/ref}
\end{subfigure}
\\
%
%
\begin{subfigure}[b]{0.1595\textwidth}
	\centering
	\includegraphics[width=28.5mm]{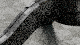}
\end{subfigure}
\begin{subfigure}[b]{0.1595\textwidth}
	\centering
	\includegraphics[width=28.5mm]{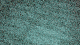}
\end{subfigure}
\begin{subfigure}[b]{0.1595\textwidth}
	\centering
	\includegraphics[width=28.5mm]{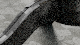}
\end{subfigure}
\begin{subfigure}[b]{0.1595\textwidth}
	\centering
	\includegraphics[width=28.5mm]{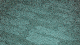}
\end{subfigure}
\begin{subfigure}[b]{0.1595\textwidth}
	\centering
	\includegraphics[width=28.5mm]{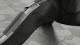}
\end{subfigure}
\begin{subfigure}[b]{0.1595\textwidth}
	\centering
	\includegraphics[width=28.5mm]{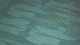}
\end{subfigure}
\\
\begin{subfigure}[b]{0.325\textwidth}
	\centering
	\includegraphics[width=58.0mm]{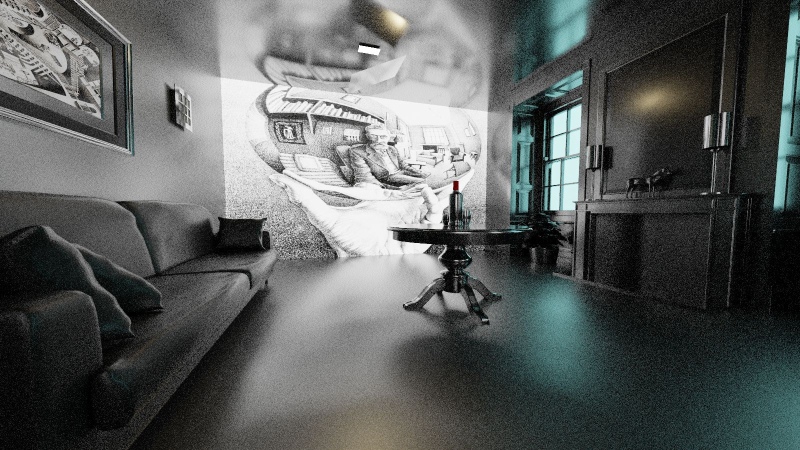}
\end{subfigure}
\begin{subfigure}[b]{0.325\textwidth}
	\centering
	\includegraphics[width=58.0mm]{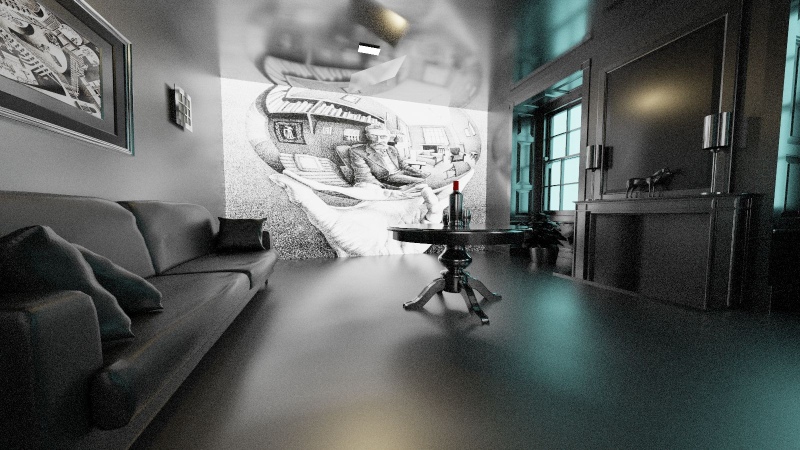}
\end{subfigure}
\begin{subfigure}[b]{0.325\textwidth}
	\centering
	\includegraphics[width=58.0mm]{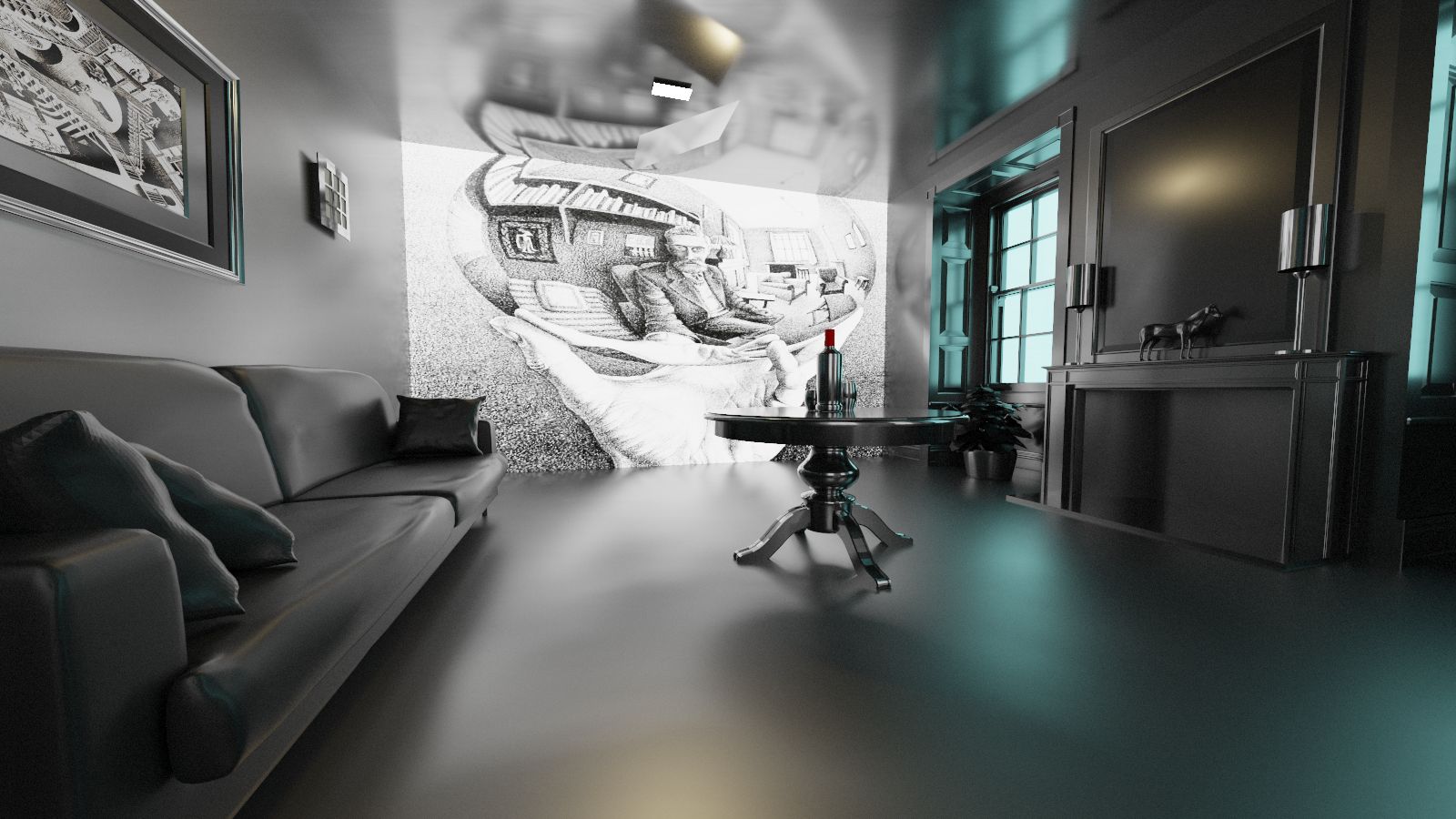}
\end{subfigure}
\\
%
%
\begin{subfigure}[b]{0.1595\textwidth}
	\centering
	\includegraphics[width=28.5mm]{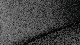}
\end{subfigure}
\begin{subfigure}[b]{0.1595\textwidth}
	\centering
	\includegraphics[width=28.5mm]{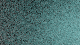}
\end{subfigure}
\begin{subfigure}[b]{0.1595\textwidth}
	\centering
	\includegraphics[width=28.5mm]{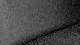}
\end{subfigure}
\begin{subfigure}[b]{0.1595\textwidth}
	\centering
	\includegraphics[width=28.5mm]{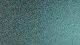}
\end{subfigure}
\begin{subfigure}[b]{0.1595\textwidth}
	\centering
	\includegraphics[width=28.5mm]{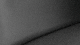}
\end{subfigure}
\begin{subfigure}[b]{0.1595\textwidth}
	\centering
	\includegraphics[width=28.5mm]{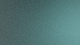}
\end{subfigure}
\\
\begin{subfigure}[b]{0.325\textwidth}
	\centering
	\includegraphics[width=58.0mm]{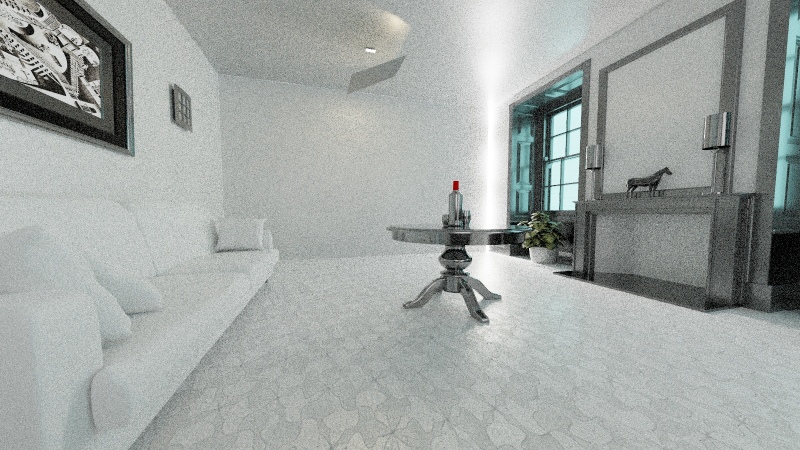}
	\caption{MMLT}
\end{subfigure}
\begin{subfigure}[b]{0.325\textwidth}
	\centering
	\includegraphics[width=58.0mm]{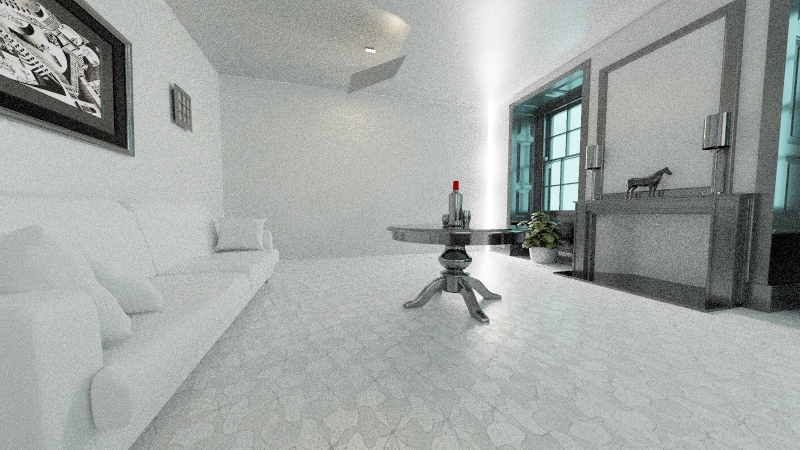}
	\caption{CMLT}
\end{subfigure}
\begin{subfigure}[b]{0.325\textwidth}
	\centering
	\includegraphics[width=58.0mm]{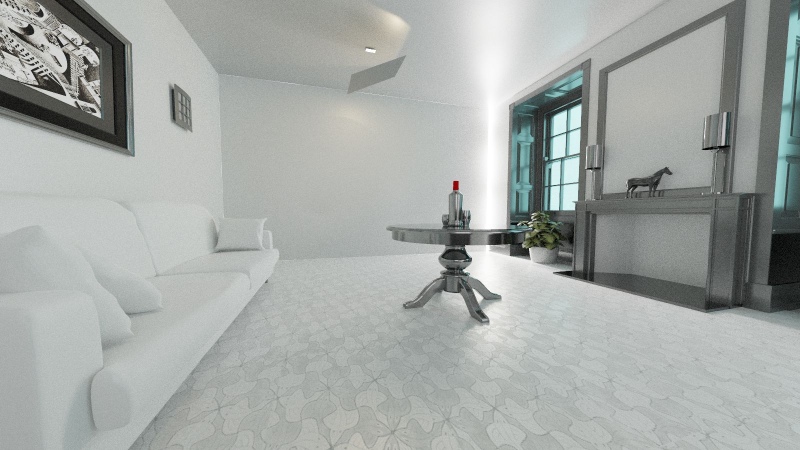}
	\caption{Reference}
\end{subfigure}
\\
%
%
\begin{subfigure}[b]{0.1595\textwidth}
	\centering
	\includegraphics[width=28.5mm]{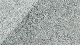}
\end{subfigure}
\begin{subfigure}[b]{0.1595\textwidth}
	\centering
	\includegraphics[width=28.5mm]{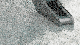}
\end{subfigure}
\begin{subfigure}[b]{0.1595\textwidth}
	\centering
	\includegraphics[width=28.5mm]{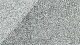}
\end{subfigure}
\begin{subfigure}[b]{0.1595\textwidth}
	\centering
	\includegraphics[width=28.5mm]{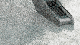}
\end{subfigure}
\begin{subfigure}[b]{0.1595\textwidth}
	\centering
	\includegraphics[width=28.5mm]{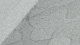}
\end{subfigure}
\begin{subfigure}[b]{0.1595\textwidth}
	\centering
	\includegraphics[width=28.5mm]{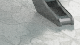}
\end{subfigure}

	\caption{Equal-time comparisons of our CMLT and MMLT on four scenes testing different transport phenomena.}
	\label{CMLT-same-time}
\end{figure*}

Figure~\ref{CMLT-same-time} shows equal-time comparisons of MMLT and CMLT on all scenes.
Except for the last row, both the MMLT and CMLT renders were generated using 256 spp, taking roughly the same computation time, whereas the reference images have been rendered with bidirectional path tracing using $16 \cdot 10^3$ spp. The images in the last row used 512 spp for MMLT and CMLT, and $64 \cdot 10^3$ spp for the reference image.

CMLT produces considerably less noise on all test scenes. In particular, it is very effective in cases of complex glossy reflections and reflections of caustics, where there is no clear winner among all bidirectional sampling techniques.

\begin{figure}
	\centering
	\begin{subfigure}[b]{0.235\textwidth}
		\centering
		\includegraphics[width=42.0mm]{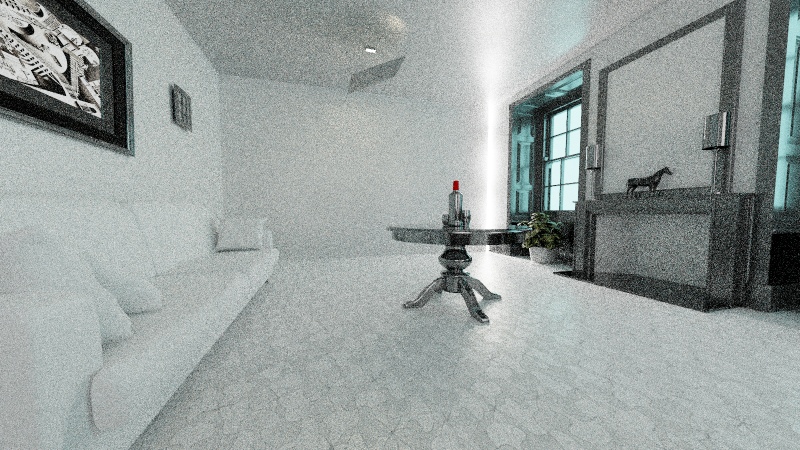}
		\caption{PSSMLT - 128 spp}
	\end{subfigure}
	\begin{subfigure}[b]{0.235\textwidth}
		\centering
		\includegraphics[width=42.0mm]{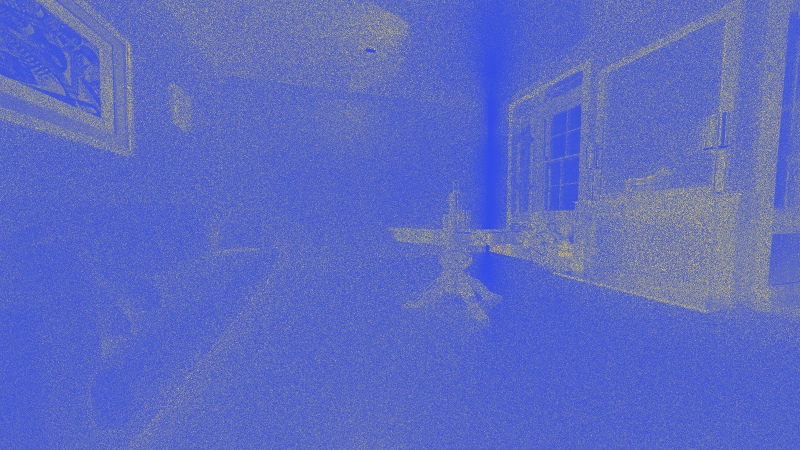}
		\caption{RMSE: 0.1977}
	\end{subfigure} \\
	\begin{subfigure}[b]{0.235\textwidth}
		\centering
		\includegraphics[width=42.0mm]{results/escher-room-ajar/mmlt-256}
		\caption{MMLT - 256 spp}
	\end{subfigure}
	\begin{subfigure}[b]{0.235\textwidth}
		\centering
		\includegraphics[width=42.0mm]{results/escher-room-ajar/mmlt-256-diff}
		\caption{RMSE: 0.1808}
	\end{subfigure} \\
	\begin{subfigure}[b]{0.235\textwidth}
		\centering
		\includegraphics[width=42.0mm]{results/escher-room-ajar/cmlt-256}
		\caption{CMLT - 256 spp}
	\end{subfigure} 
	\begin{subfigure}[b]{0.235\textwidth}
		\centering
		\includegraphics[width=42.0mm]{results/escher-room-ajar/cmlt-256-diff}
		\caption{RMSE: 0.1546}
	\end{subfigure}
	
	\caption{RMSE comparison of PSSMLT (top), MMLT (middle) and CMLT (bottom) at equal computation time.}
	\label{CMLT-convergence-2}
\end{figure}

Figure~\ref{CMLT-convergence} shows the convergence of MMLT and CMLT on the Salle de bain scene. Notice how MMLT needs more than twice as many samples as CMLT to get approximately the same RMSE.
In the early stages, MMLT is not capable of finding many important light paths, leading to an apparently darker image (due to energy being concentrated on a subset of the pixels); the difference vanishes at higher sampling rates.
Figure~\ref{CMLT-convergence-2} shows a similar graph comparing also to PSSMLT.
Since each PSSMLT sample requires both more shadow rays and BSDF evaluations, in our implementations PSSMLT can perform roughly one half the mutations as CMLT in the same time.

\begin{figure}
	\centering
	
	\begin{subfigure}[b]{0.15\textwidth}
		\centering
		\includegraphics[width=27.0mm]{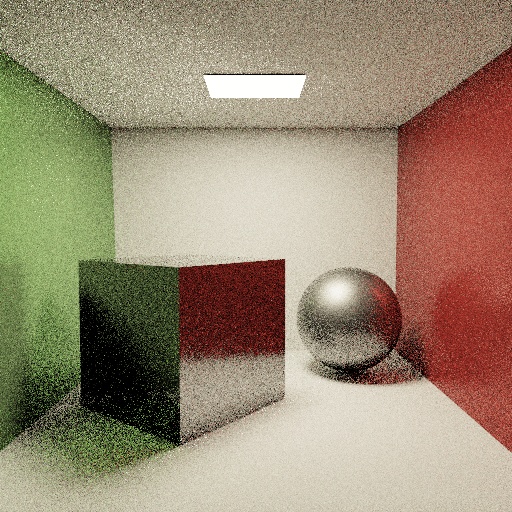}
	\end{subfigure}
	\begin{subfigure}[b]{0.15\textwidth}
		\centering
		\includegraphics[width=27.0mm]{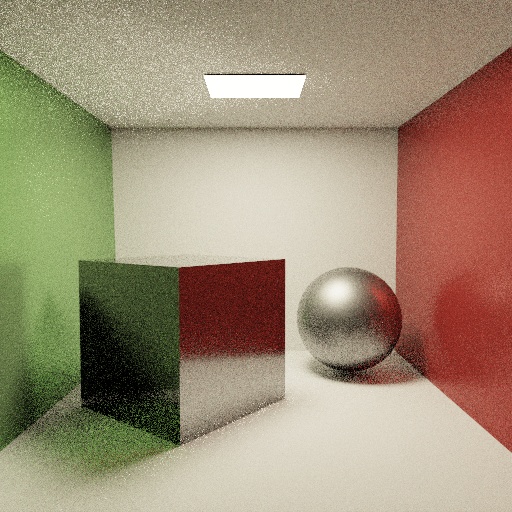}
	\end{subfigure} 
	\begin{subfigure}[b]{0.15\textwidth}
		\centering
		\includegraphics[width=27.0mm]{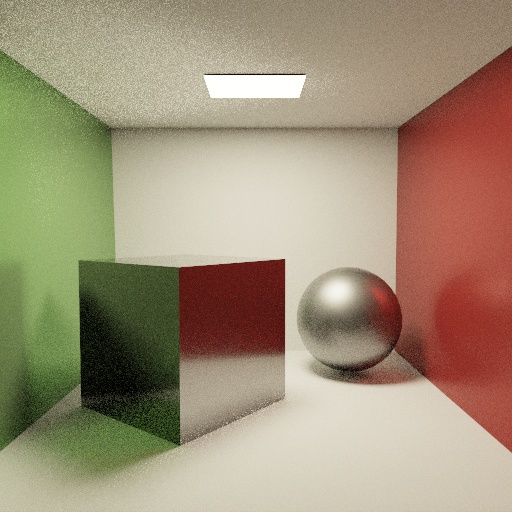}
	\end{subfigure} 
	\\
	\begin{subfigure}[b]{0.15\textwidth}
		\centering
		\includegraphics[width=27.0mm]{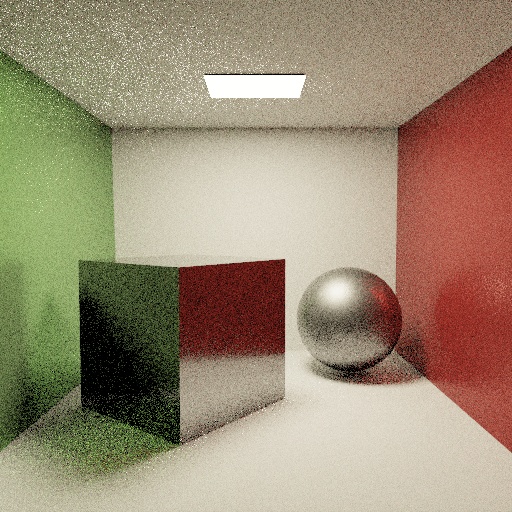}
		\caption{32 spp}
	\end{subfigure}
	\begin{subfigure}[b]{0.15\textwidth}
		\centering
		\includegraphics[width=27.0mm]{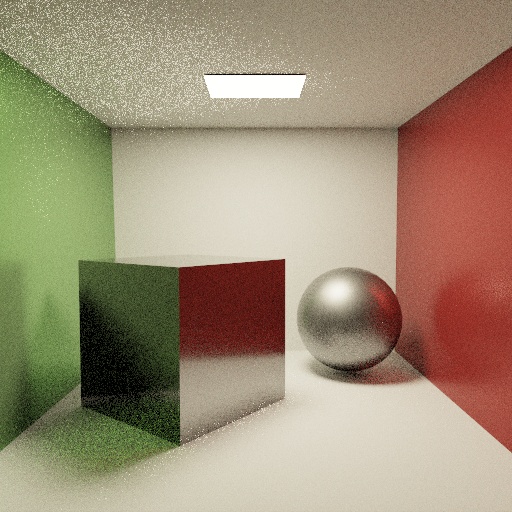}
		\caption{128 spp}
	\end{subfigure} 
	\begin{subfigure}[b]{0.15\textwidth}
		\centering
		\includegraphics[width=27.0mm]{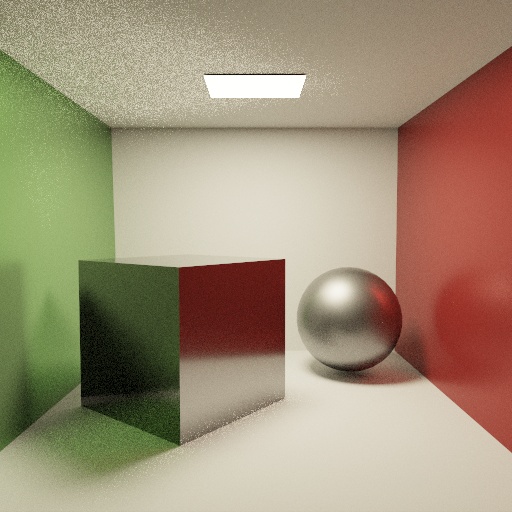}
		\caption{512 spp}
	\end{subfigure}
	
	\caption{Parallel CMLT convergence using respectively 32K (top row) and 256K chains (bottom row). From left to right: 32, 128 and 512 samples per pixel. }
	\label{CMLT-stratification}
\end{figure}

Finally, Figure~\ref{CMLT-stratification} shows the effect of varying the number of chains run in parallel, trading it against chain length to keep the total number of samples fixed. The images in the top row are obtained running $32 \cdot 10^3$ chains in parallel, whereas the ones in the bottom row are obtained using $256 \cdot 10^3$ chains. It can be seen that using more, shorter chains generally improves stratification and widely reduces the spotty appearance typical of Metropolis autocorrelation.
The exception is the caustic on the ceiling that benefits from the higher adaptation of the longer chains.
Note that the additional stratification is similar to the one obtained by ERPT \cite{Cline:2005}, which however was using a different, per-pixel chain distribution strategy (as opposed to our global resampling stage), and was not specifically targeted at introducing massive parallelism.
While Cline et al \shortcite{Cline:2005} discussed only the stratification benefits, we believe it is worth documenting what seems an intrinsic tradeoff between local exploration and better stratification: running more, shorter chains generally helps image stratification, while necessarily losing some exploration capabilities in narrow regions of path space.

In all cases, for CMLT we used one chart swap proposal every 16 mutations, resulting in negligible overhead.

\ifnum 0 = 0

\begin{figure*}
	\centering

\begin{subfigure}[b]{0.195\textwidth}
	\centering
	\includegraphics[width=34.5mm]{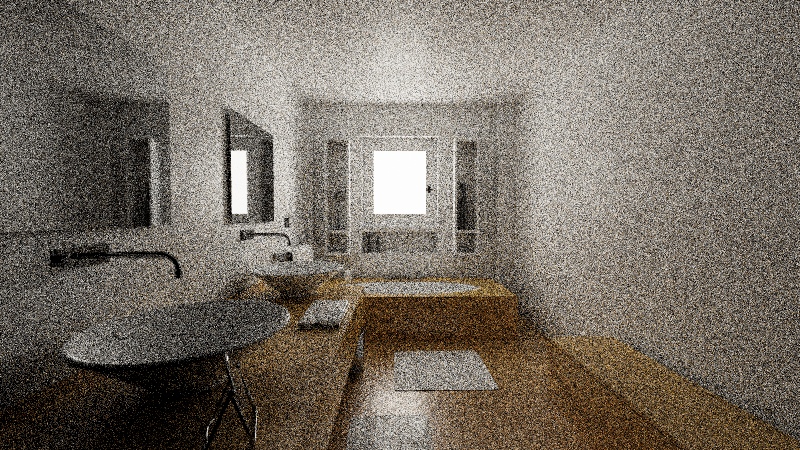}
	\caption{MMLT - 4 spp}
\end{subfigure}
\begin{subfigure}[b]{0.195\textwidth}
	\centering
	\includegraphics[width=34.5mm]{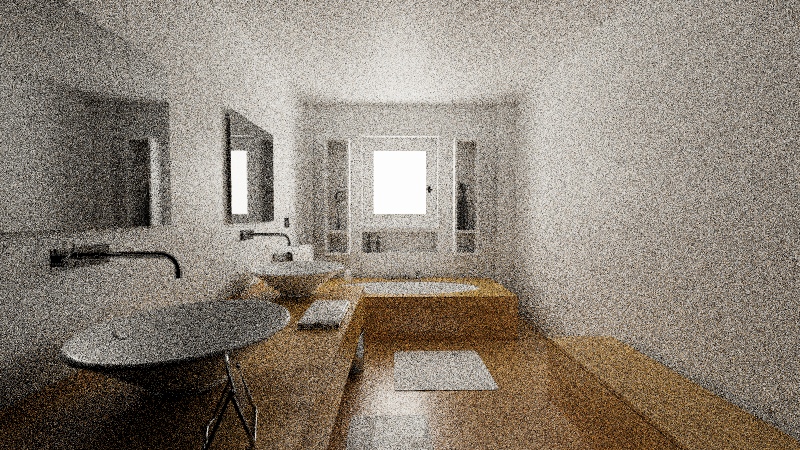}
	\caption{MMLT - 8 spp}
\end{subfigure}
\begin{subfigure}[b]{0.195\textwidth}
	\centering
	\includegraphics[width=34.5mm]{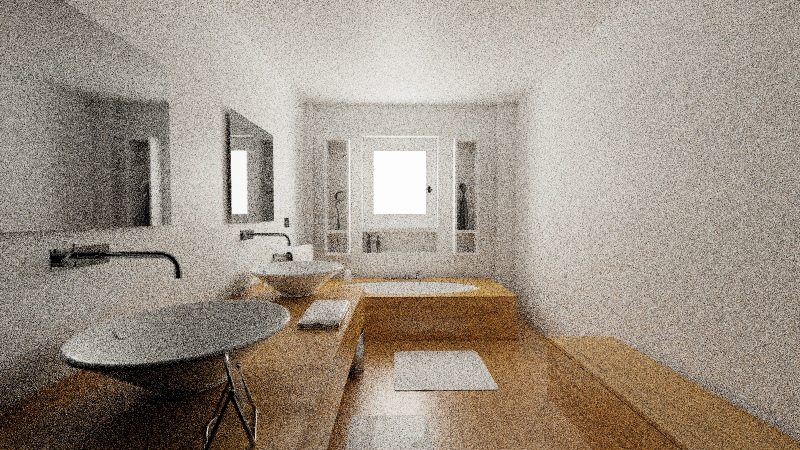}
	\caption{MMLT - 16 spp}
\end{subfigure}
\begin{subfigure}[b]{0.195\textwidth}
	\centering
	\includegraphics[width=34.5mm]{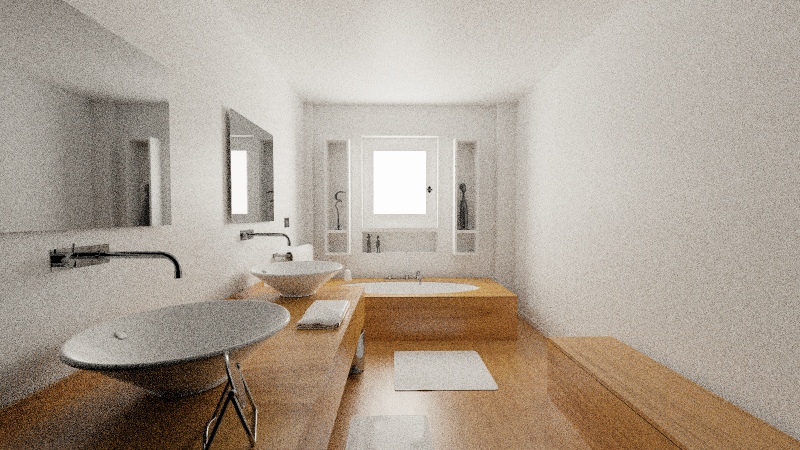}
	\caption{MMLT - 64 spp}
\end{subfigure}
\begin{subfigure}[b]{0.195\textwidth}
	\centering
	\includegraphics[width=34.5mm]{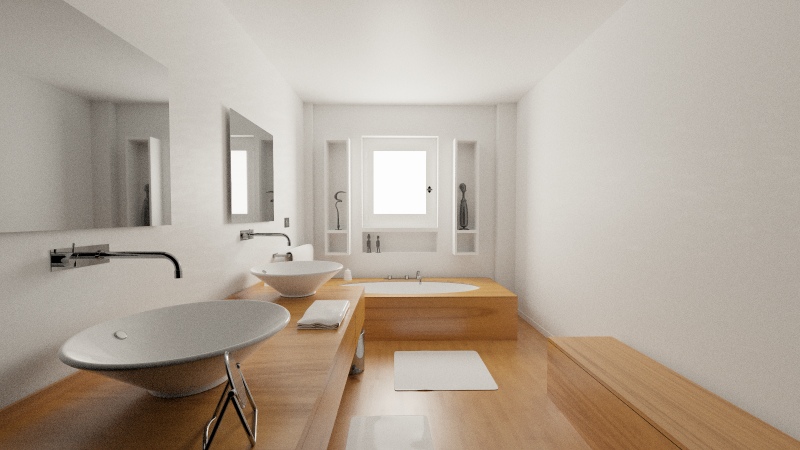}
	\caption{MMLT - 1024 spp}
\end{subfigure}
\\
\begin{subfigure}[b]{0.195\textwidth}
	\centering
	\includegraphics[width=34.5mm]{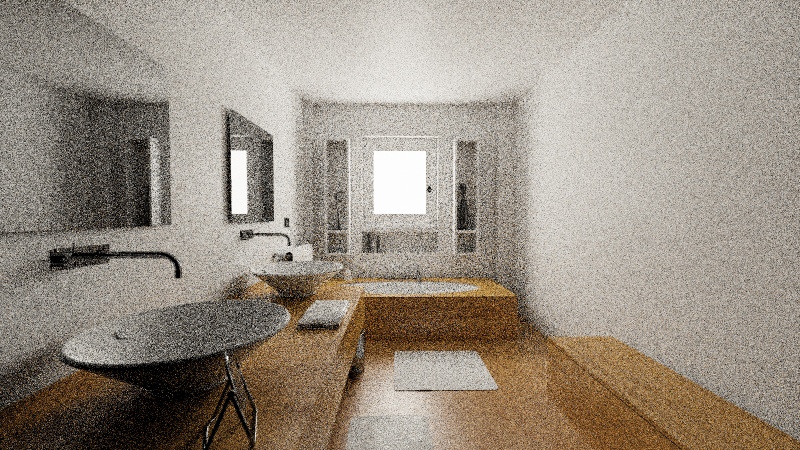}
	\caption{CMLT - 4 spp}
\end{subfigure}
\begin{subfigure}[b]{0.195\textwidth}
	\centering
	\includegraphics[width=34.5mm]{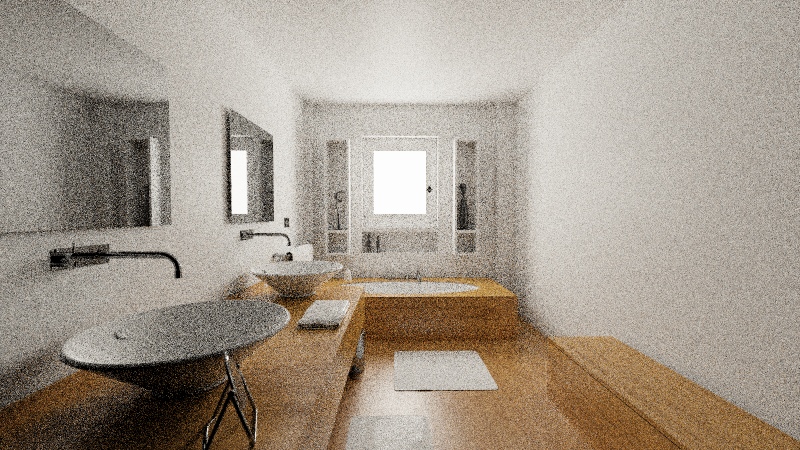}
	\caption{CMLT - 8 spp}
\end{subfigure}
\begin{subfigure}[b]{0.19\textwidth}
	\centering
	\includegraphics[width=34.5mm]{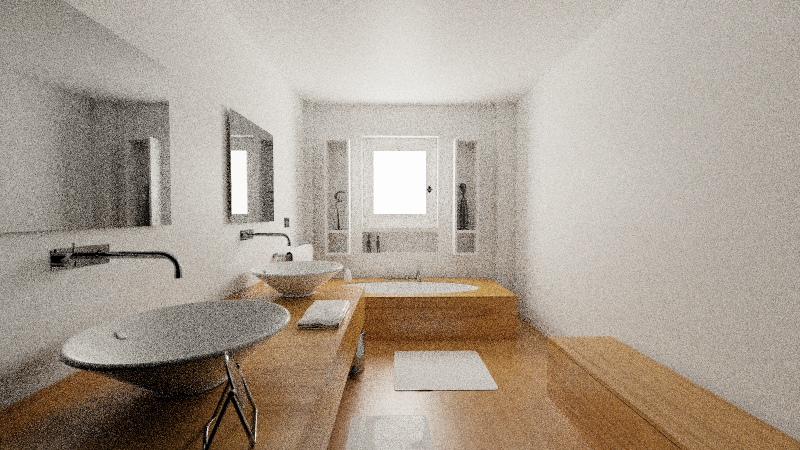}
	\caption{CMLT - 16 spp}
\end{subfigure}
\begin{subfigure}[b]{0.195\textwidth}
	\centering
	\includegraphics[width=34.5mm]{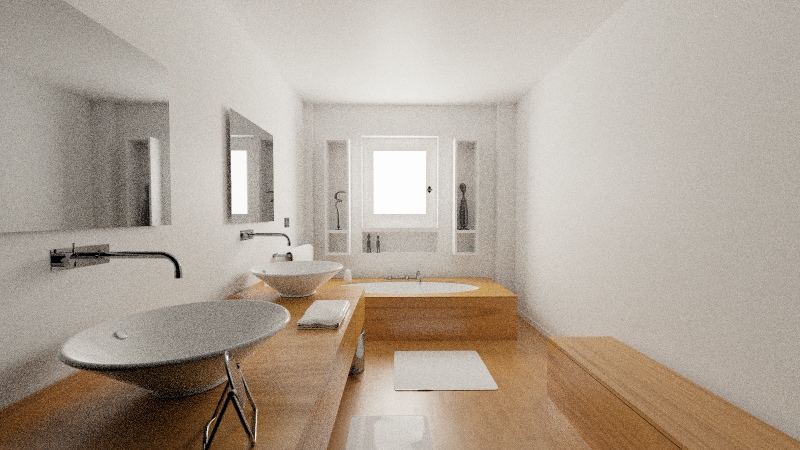}
	\caption{CMLT - 64 spp}
\end{subfigure}
\begin{subfigure}[b]{0.195\textwidth}
	\centering
	\includegraphics[width=34.5mm]{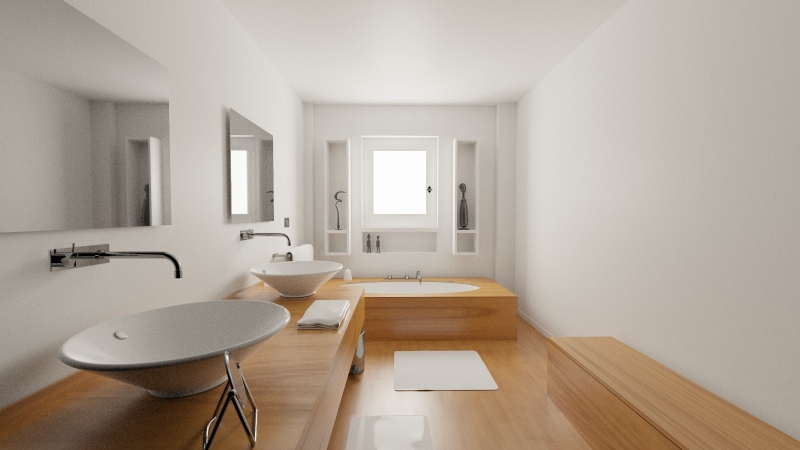}
	\caption{CMLT - 1024 spp}
\end{subfigure}
\\
\begin{subfigure}[b]{0.195\textwidth}
	\centering
	\includegraphics[width=34.5mm]{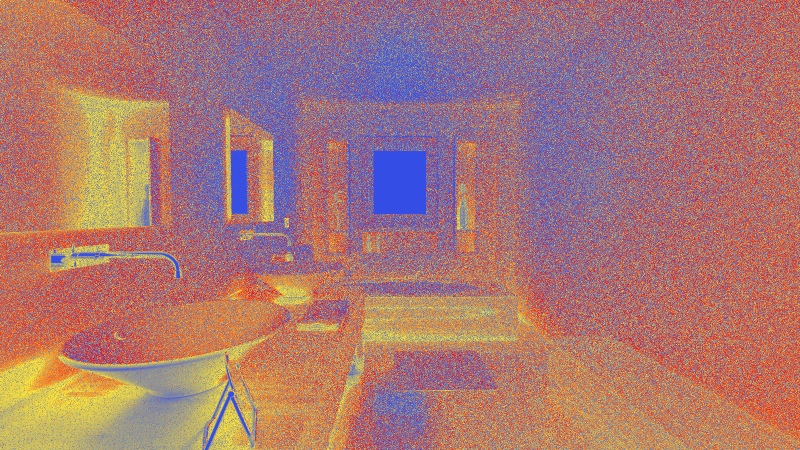}
	\caption{RMSE: 0.6490}
\end{subfigure}
\begin{subfigure}[b]{0.195\textwidth}
	\centering
	\includegraphics[width=34.5mm]{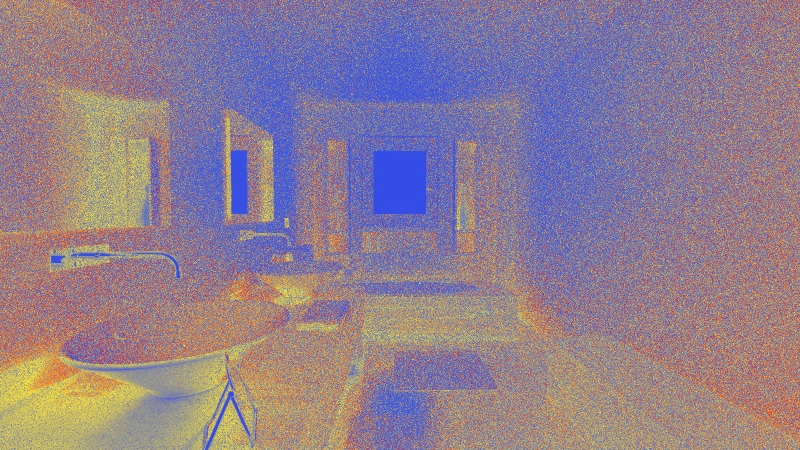}
	\caption{RMSE: 0.4914}
\end{subfigure}
\begin{subfigure}[b]{0.195\textwidth}
	\centering
	\includegraphics[width=34.5mm]{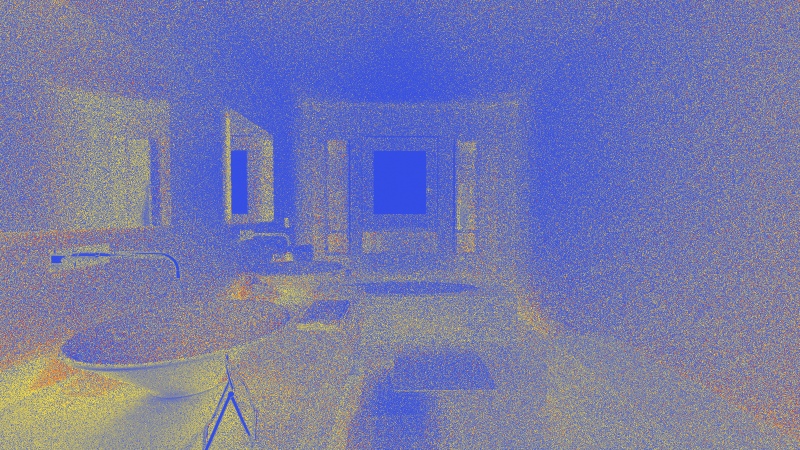}
	\caption{RMSE: 0.3439}
\end{subfigure}
\begin{subfigure}[b]{0.195\textwidth}
	\centering
	\includegraphics[width=34.5mm]{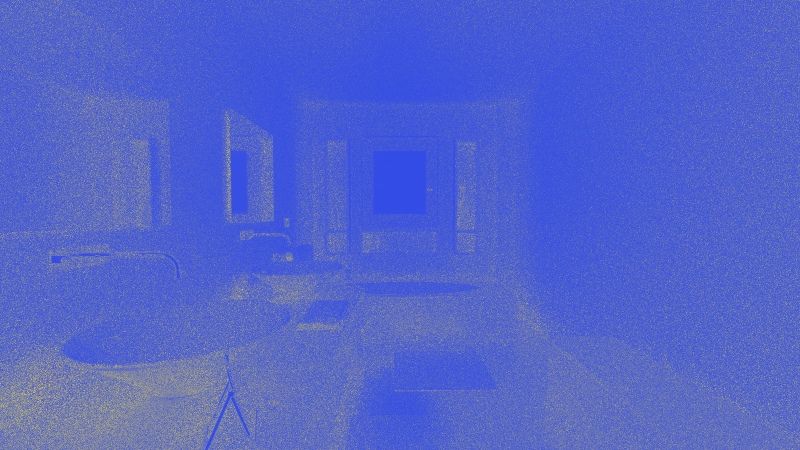}
	\caption{RMSE: 0.1552}
\end{subfigure}
\begin{subfigure}[b]{0.195\textwidth}
	\centering
	\includegraphics[width=34.5mm]{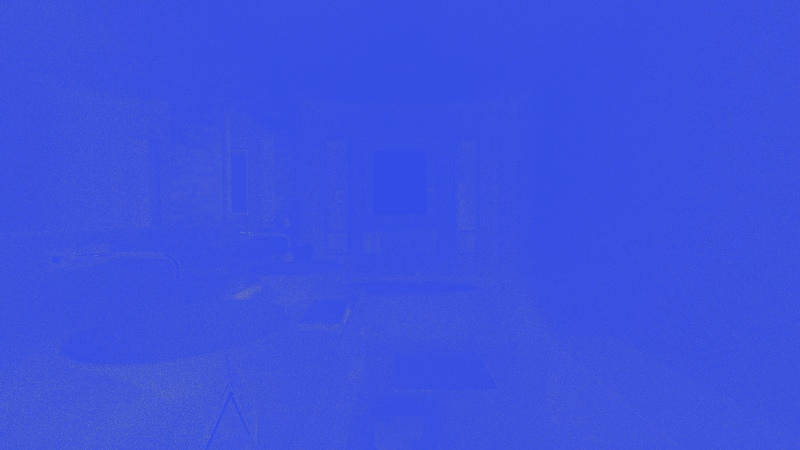}
	\caption{RMSE: 0.0373}
\end{subfigure}
\\
\begin{subfigure}[b]{0.195\textwidth}
	\centering
	\includegraphics[width=34.5mm]{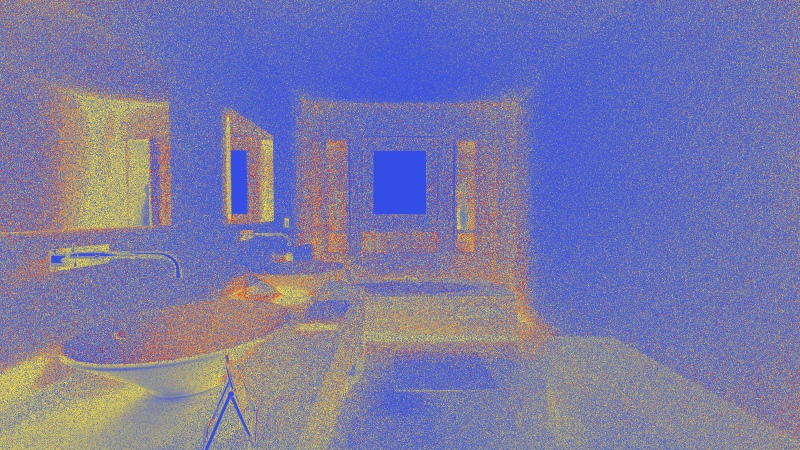}
	\caption{RMSE: 0.3697}
\end{subfigure}
\begin{subfigure}[b]{0.195\textwidth}
	\centering
	\includegraphics[width=34.5mm]{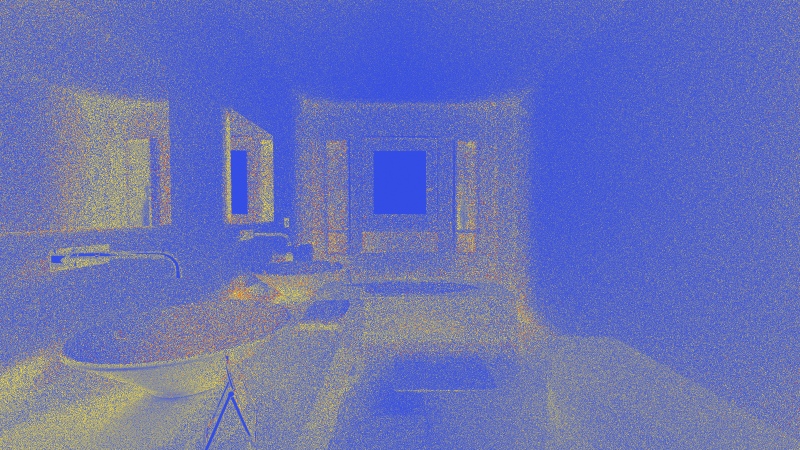}
	\caption{RMSE: 0.2683}
\end{subfigure}
\begin{subfigure}[b]{0.195\textwidth}
	\centering
	\includegraphics[width=34.5mm]{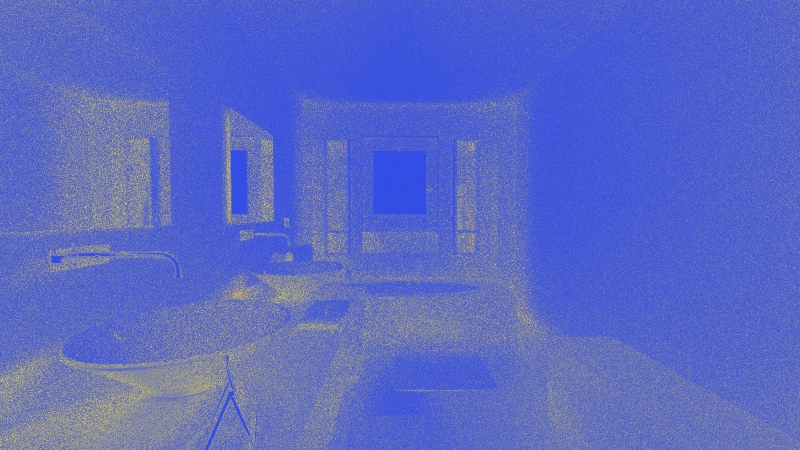}
	\caption{RMSE: 0.1913}
\end{subfigure}
\begin{subfigure}[b]{0.195\textwidth}
	\centering
	\includegraphics[width=34.5mm]{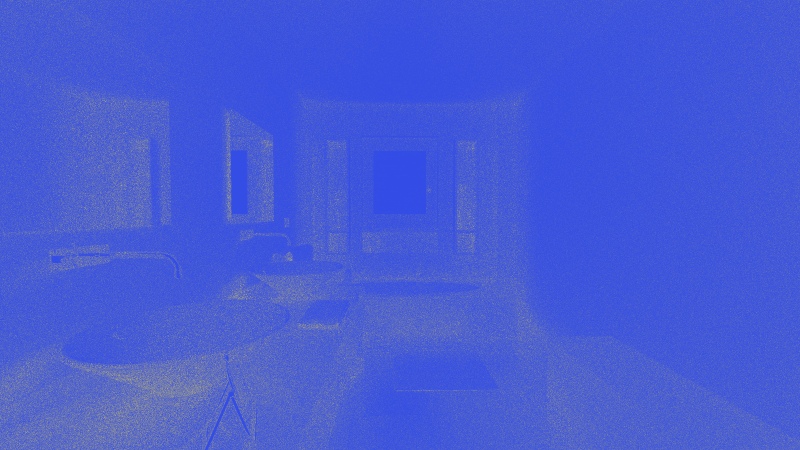}
	\caption{RMSE: 0.0964}
\end{subfigure}
\begin{subfigure}[b]{0.195\textwidth}
	\centering
	\includegraphics[width=34.5mm]{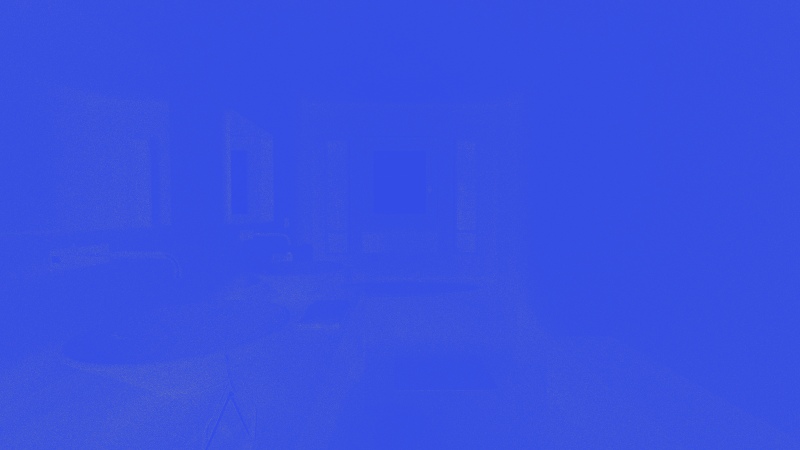}
	\caption{RMSE: 0.0251}
\end{subfigure}

\caption{RMSE comparison of CMLT (bottom) and MMLT (top) at 4, 8, 16, 64 and 1024 spp.}
\label{CMLT-convergence}
\end{figure*}

\else

\begin{figure*}
	\centering
	
\begin{subfigure}[b]{0.195\textwidth}
	\centering
	\includegraphics[width=34.5mm]{results/escher-room-ajar/mmlt-64}
	\caption{MMLT - 64 spp}
\end{subfigure}
\begin{subfigure}[b]{0.195\textwidth}
	\centering
	\includegraphics[width=34.5mm]{results/escher-room-ajar/mmlt-128}
	\caption{MMLT - 128 spp}
\end{subfigure}
\begin{subfigure}[b]{0.195\textwidth}
	\centering
	\includegraphics[width=34.5mm]{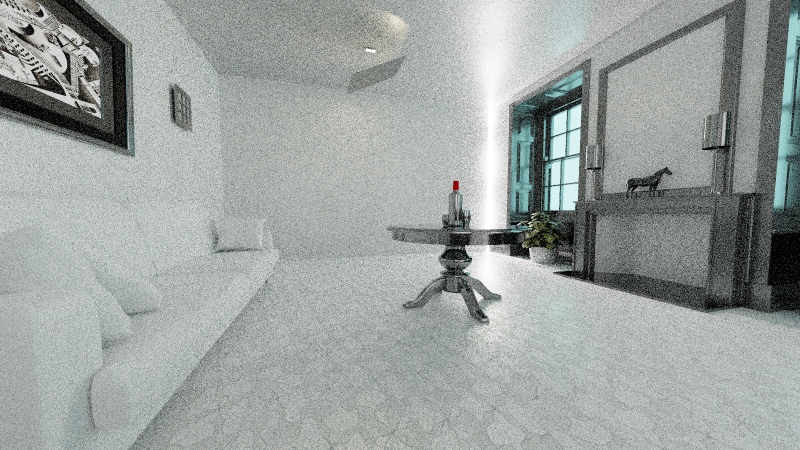}
	\caption{MMLT - 256 spp}
\end{subfigure}
\begin{subfigure}[b]{0.195\textwidth}
	\centering
	\includegraphics[width=34.5mm]{results/escher-room-ajar/mmlt-512}
	\caption{MMLT - 512 spp}
\end{subfigure}
\begin{subfigure}[b]{0.195\textwidth}
	\centering
	\includegraphics[width=34.5mm]{results/escher-room-ajar/mmlt-1024}
	\caption{MMLT - 1024 spp}
\end{subfigure}
	\\
\begin{subfigure}[b]{0.195\textwidth}
	\centering
	\includegraphics[width=34.5mm]{results/escher-room-ajar/cmlt-64}
	\caption{CMLT - 64 spp}
\end{subfigure}
\begin{subfigure}[b]{0.195\textwidth}
	\centering
	\includegraphics[width=34.5mm]{results/escher-room-ajar/cmlt-128}
	\caption{CMLT - 128 spp}
\end{subfigure}
\begin{subfigure}[b]{0.19\textwidth}
	\centering
	\includegraphics[width=34.5mm]{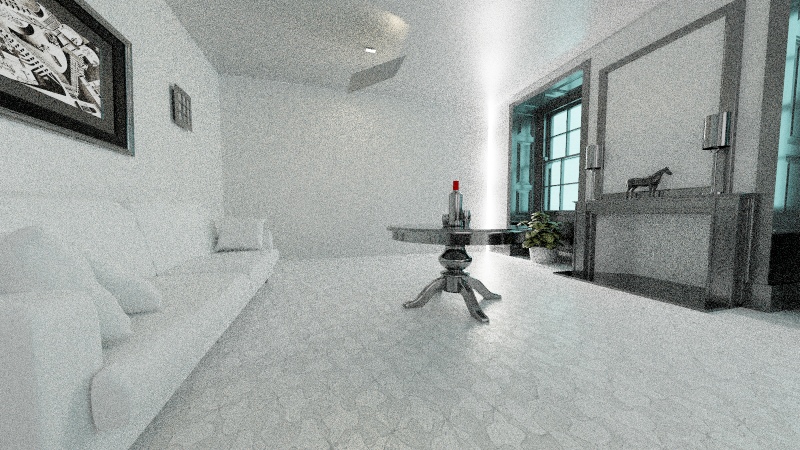}
	\caption{CMLT - 256 spp}
\end{subfigure}
\begin{subfigure}[b]{0.195\textwidth}
	\centering
	\includegraphics[width=34.5mm]{results/escher-room-ajar/cmlt-512}
	\caption{CMLT - 512 spp}
\end{subfigure}
\begin{subfigure}[b]{0.195\textwidth}
	\centering
	\includegraphics[width=34.5mm]{results/escher-room-ajar/cmlt-1024}
	\caption{CMLT - 1024 spp}
\end{subfigure}
\\
\begin{subfigure}[b]{0.195\textwidth}
	\centering
	\includegraphics[width=34.5mm]{results/escher-room-ajar/mmlt-64-diff}
	\caption{RMSE: 0.4158}
\end{subfigure}
\begin{subfigure}[b]{0.195\textwidth}
	\centering
	\includegraphics[width=34.5mm]{results/escher-room-ajar/mmlt-128-diff}
	\caption{RMSE: 0.2730}
\end{subfigure}
\begin{subfigure}[b]{0.195\textwidth}
	\centering
	\includegraphics[width=34.5mm]{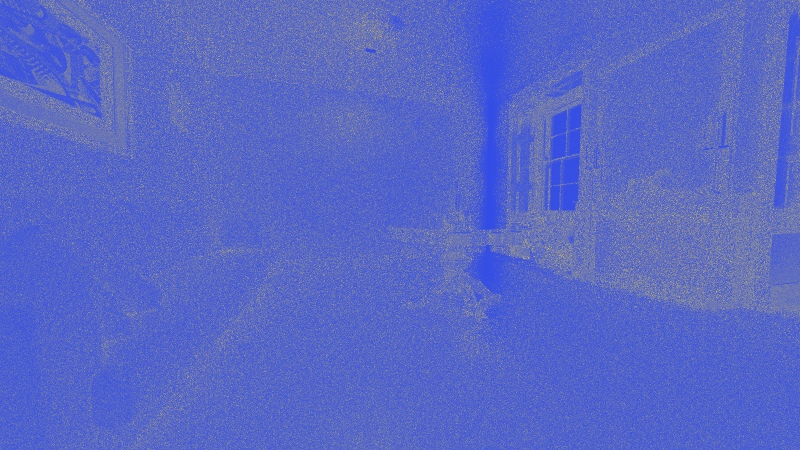}
	\caption{RMSE: 0.1808}
\end{subfigure}
\begin{subfigure}[b]{0.195\textwidth}
	\centering
	\includegraphics[width=34.5mm]{results/escher-room-ajar/mmlt-512-diff}
	\caption{RMSE: 0.1206}
\end{subfigure}
\begin{subfigure}[b]{0.195\textwidth}
	\centering
	\includegraphics[width=34.5mm]{results/escher-room-ajar/mmlt-1024-diff}
	\caption{RMSE: 0.0811}
\end{subfigure}
\\
\begin{subfigure}[b]{0.195\textwidth}
	\centering
	\includegraphics[width=34.5mm]{results/escher-room-ajar/cmlt-64-diff}
	\caption{RMSE: 0.3143}
\end{subfigure}
\begin{subfigure}[b]{0.195\textwidth}
	\centering
	\includegraphics[width=34.5mm]{results/escher-room-ajar/cmlt-128-diff}
	\caption{RMSE: 0.2224}
\end{subfigure}
\begin{subfigure}[b]{0.195\textwidth}
	\centering
	\includegraphics[width=34.5mm]{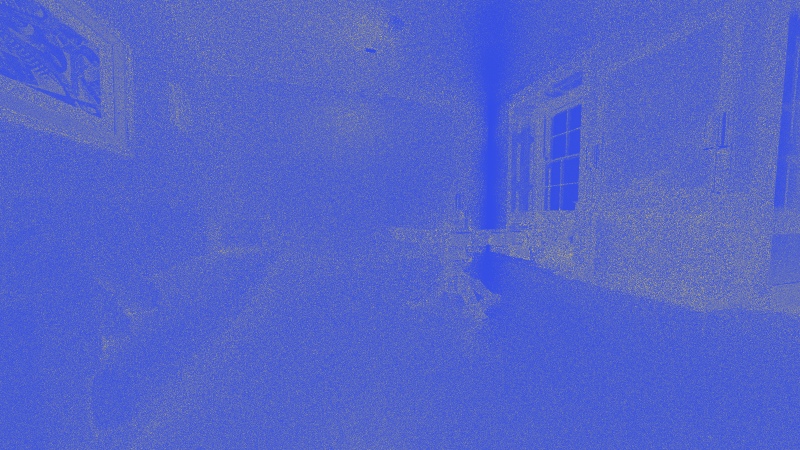}
	\caption{RMSE: 0.1546}
\end{subfigure}
\begin{subfigure}[b]{0.195\textwidth}
	\centering
	\includegraphics[width=34.5mm]{results/escher-room-ajar/cmlt-512-diff}
	\caption{RMSE: 0.1060}
\end{subfigure}
\begin{subfigure}[b]{0.195\textwidth}
	\centering
	\includegraphics[width=34.5mm]{results/escher-room-ajar/cmlt-1024-diff}
	\caption{RMSE: 0.0721}
\end{subfigure}

	\caption{RMSE comparison of CMLT (bottom) and MMLT (top) at 64, 128, 256, 512 and 1024 spp.}
	\label{CMLT-convergence}
\end{figure*}

\fi

\subsection{Performance analysis}

\begin{figure}
	\fbox{\includegraphics[width=82.0mm]{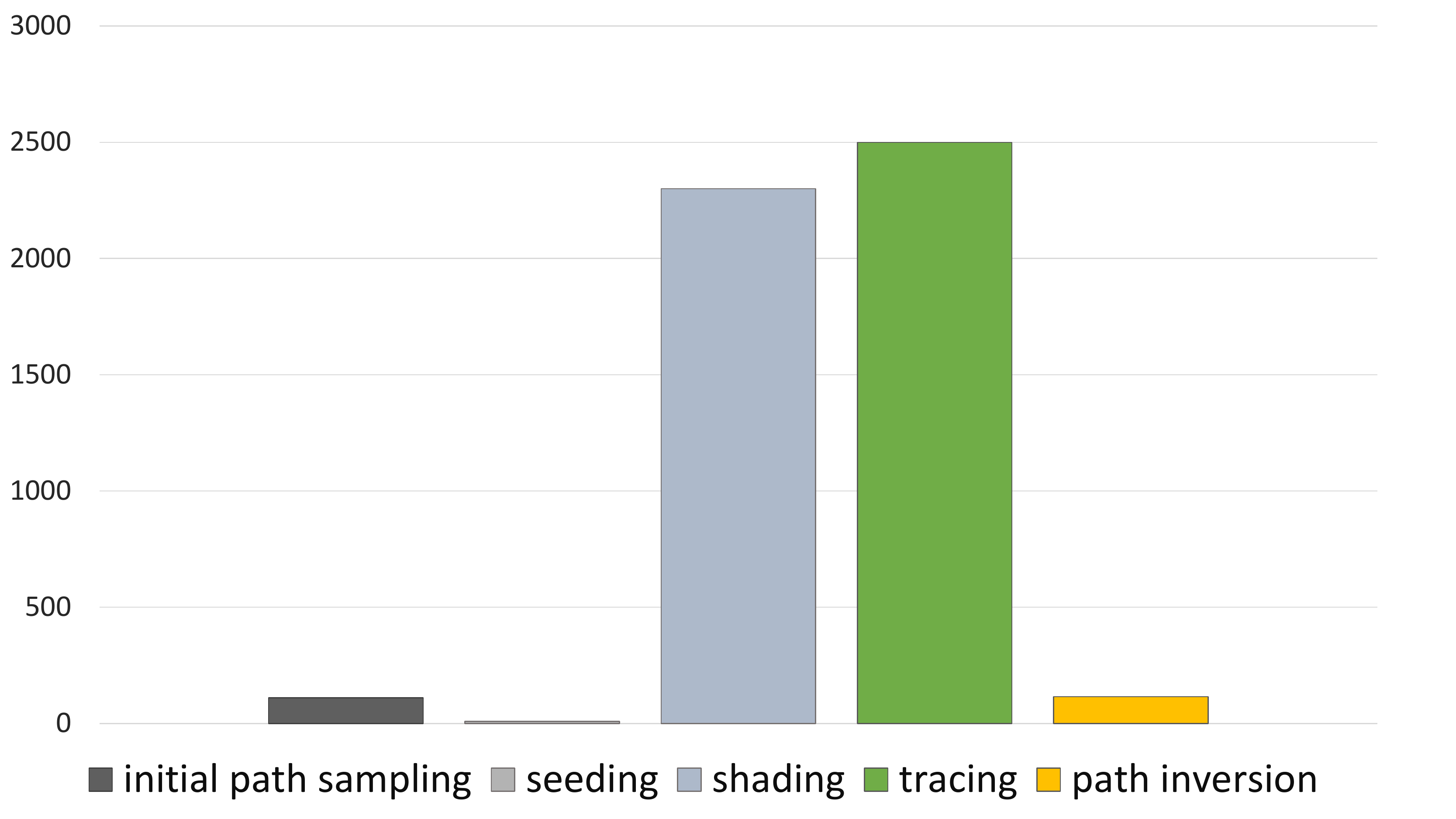}}
	\caption{Performance breakdown for running 256K chains of length 350 (equivalent to about 64 spp at a resolution of $1600 \times 900$). All timings are in milliseconds. }
	\label{PerfBreakdown}
\end{figure}

On our system, the 1024 spp CMLT and MMLT images take roughly 80s to render at a resolution of $1600 \times 900$ using $256 \cdot 10^3$ chains.
Figure~\ref{PerfBreakdown} shows a performance breakdown on Salle de bain: roughly 50\% of the time is spent in ray tracing, with shading taking 45\%, and the initial path sampling and path inversion taking roughly 2.5\% each.

If we substantially reduce the number of chains we start to notice a slowdown due to underutilization of the hardware resources, mostly caused by insufficient parallelism in the late stages of the bidirectional path tracing pipeline needed to process longer than average paths.
This could likely be mitigated by better scheduling policies, for example not requiring all chains to be processed in sync (currently we finish applying a mutation to all paths before starting to process the next).

\section{Discussion}

We proposed a novel family of MCMC algorithms that use sampling charts to extend the sampling domain and allow better exploration.
We applied the new scheme to propose a new type of light transport simulation algorithms that bridge primary sample space and path space MLT.

We also showed that the new algorithms arising from this framework require to implement only a new set of relatively cheap mutations that can be constructed using simple, stochastic right inverses of the path sampling functions: particularly, the fact our framework requires only such type of probabilistic inversion is what makes the algorithm practical, as classical BSDF inversion with layered material models is generally impossible.
We believe this to be a major strength of our work.

We implemented both the old and new methods exposing massive parallelism at all levels, and showed how increasing the number of chains that run in parallel can increase stratification.

Finally, we suggested a novel, simpler method to integrate path density estimation into MCMC light transport algorithms as a mechanism to craft independent proposals.

\subsection{Future work}

There are multiple avenues in which this work could be extended.
The first is testing all possible variants of our new algorithmic family more thoroughly.
In such a context, it will be particularly interesting to test the combination with the original path space MLT mutations, which might provide some advantages in regions with complex visibility.
Similarly, it would be interesting to test the new technique for including path density estimation as an independence sampler.

Another potential venue is considering \emph{dimension jumps} to switch between the charts underlying different path spaces $\Omega_k$ and $\Omega_{k'}$. This could be achieved using the \emph{Metropolis-Hastings-Green with Jacobians} algorithm as described by Geyer \shortcite{Geyer:2011}.

Finally, it would be interesting to integrate half vector space light transport \cite{Kaplanyan:2014:HSLT,Hanika:2015:IHLST} as yet another path space chart.

\paragraph{{\bf Aknowledgements}}
We would like to thank Cem Cebenoyan at NVIDIA for constantly supporting our work; Luca Fascione and Marc Droske at Weta Digital for early reviews and continuous feedback; Matthias Raab at NVIDIA for helping us with modern layered material sampling methods; Nicholas Hull and Nir Benty at NVIDIA for their precious help with the setup and import of the original Gray \& White Room and Salle de bain scenes and Thomas Iuliano for providing beautiful artwork that ought to be included in this paper, and was not for mere lack of time.
Finally, we would like to thank the anonymous SIGGRAPH reviewers, particularly \#30, for their detailed comments which led to significant improvements in the exposition of the paper.

\section{Appendix}

We here describe how to invert the sampling functions for typical BSDF layers as needed to implement chart swaps.
The key insight is that most common BSDF sampling methods can be seen as bijective functions $S(\omega_i)$ from the unit square to the hemisphere of directions:
\begin{eqnarray}
S(\omega_i):[0,1]^2 &\rightarrow& H \\
(u,v) &\mapsto& \omega_o \nonumber
\end{eqnarray}
where the notation $S(\omega_i)$ denotes the potential dependence on the incident direction $\omega_i$.
Hence, in order to perform BSDF inversion, we need to simply compute the inverse $S^\leftarrow(\omega_i): H \rightarrow [0,1]^2$.

\subsection{Lambertian distribution}

Lambertian BSDFs are typically importance sampled using the mapping:
\begin{equation}
S: (u,v) \mapsto (\theta,\phi) = (acos(\sqrt{v}), u \cdot 2 \pi)
\end{equation}
where $(\theta, \phi)$ represent spherical coordinates relative to the surface normal.
Inverting this mapping can hence be done very easily:
\begin{equation}
S^\leftarrow: (\theta,\phi) \mapsto (u,v) = \left(
\frac{\phi}{2 \pi},
cos^2(\theta)
\right)
\end{equation}

\subsection{GGX distribution}

Sampling the GGX distribution is slightly more involved as it is the composition of two functions: $S(\omega_i) = R(\omega_i) \circ F_m$, where the function $F_m:[0,1]^2 \rightarrow H$ samples a microfacet according to the roughness parameter $m$, and $R(\omega_i) : H \rightarrow H$ returns the input direction $\omega_i$ reflected about the sampled microfacet normal.
Its inverse can hence be obtained as $S^\leftarrow(\omega_i) = F_m^\leftarrow \circ R^\leftarrow(\omega_i)$.

Finding the microfacet normal given the incident and outgoing directions $\omega_i$ and $\omega_o$ is trivial, as the normal can be simply computed using the half vector formula:
\begin{eqnarray}
R^\leftarrow(\omega_i) : H &\rightarrow& H \\
\omega_o &\mapsto& \frac{\omega_i + \omega_o}{|\omega_i + \omega_o|}. \nonumber
\end{eqnarray}

The forward mapping for sampling a microfacet is instead given by the following expression:
\begin{equation}
F_m: (u,v) \mapsto (\theta,\phi) = \left( acos\left(\frac{1}{\sqrt{1 + t(v)}}\right), u \cdot 2 \pi \right)
\end{equation}
with:
\begin{equation}
t(v) = \frac{v}{ (1 - v) \cdot m^2 }.
\end{equation}
The inverse can hence be computed as:
\begin{equation}
F_m^\leftarrow: (\theta,\phi) \mapsto (u,v) = \left(
\frac{\phi}{2 \pi},
\frac{q(\theta) }{1 + q(\theta) }
\right)
\end{equation}
with:
\begin{equation}
q(\theta) = m^2 \cdot (1 / cos^2(\theta) - 1).
\end{equation}

The composition of the two can now be obtained considering the polar coordinates $(\theta_h, \phi_h)$ of the vector:
\begin{equation}
{\bf h} = R^\leftarrow(\omega_i,\omega_o),
\end{equation}
and finally computing:
\begin{equation}
(u,v) = F_m^\leftarrow(\theta_h,\phi_h).
\end{equation}

\subsection{Specular scattering}

Specular scattering introduces singularities in the transformations $T_i$, which manifest as Dirac deltas in the respective pdfs $p_i$.
While we did not explicitly study how to handle these in this work, we believe it would be possible to include them in our chart swaps, as long as the scattering mode at specular vertices is not altered. In fact, altering the mode from specular to diffuse would simply result in a zero acceptance rate: this can be verified looking at equation (\ref{eqn:STAcceptanceRatio}), and considering the fact that the numerator, equal to the reciprocal of the density of the current (specular) pdf $p_i$, would be zero.
Conversely, if the mode was not altered, the implicit Dirac deltas in the numerator and denominator would cancel out.

\RestyleAlgo{boxruled}
\begin{algorithm}
	{
		// fill the three arrays:\\
		//   u\_init[] : primary sample space path coordinates\\
		//   st\_init[] : technique number of each path\\
		//   C\_init[] : contribution of each path\\
		(u\_init,st\_init,C\_init) $\leftarrow$ {\bf bptSamplePaths}($N_{init}$)\;
		\vspace{2mm}
		// build a cdf over the $N_{init}$ path contributions\\
		cdf[] $\leftarrow$ {\bf prefixSum}(C\_init)\;
		\vspace{2mm}
		// resample N paths based on their contribution\\
		\While{i=1 ... N}{
			seed $\leftarrow$ {\bf sampleCdf}(cdf,(i + random())/N)\;
			
			u[i] $\leftarrow$ u\_init[seed]\;
			
			st[i] $\leftarrow$ st\_init[seed]\;
			
			C[i] $\leftarrow$ C\_init[seed]\;
			\vspace{2mm}
			// retrace the bidirectional path\\
			path[i] $\leftarrow$ {\bf bptTracePath}(u[i], st[i])\;
		}
		
		// loop across the number of mutations L\\
		\While{l=1 ... L}{
			\While{i=1 ... N}{
				\eIf {selectMutation(l) == ChartSwap}
				{
					// propose a chart swap\\
					(s, t) $\leftarrow$ st[i]\;
					k $\leftarrow$ s + t - 1\;
					(s',t') $\leftarrow$ {\bf chartProposal}(k)\;
					\eIf {s' > s} {
						(u',r') $\leftarrow$ {\bf invertLightSubpath}(path[i],s,s')\;
						
						r $\leftarrow$ {\bf eyeSubpathInversionPdf}(path[i], t', t)\;
					} {
						(u',r') $\leftarrow$ {\bf invertEyeSubpath}(path[i],t,t')\;
						
						r $\leftarrow$ {\bf lightSubpathInversionPdf}(path[i], s', s)\;
					}
					
					// compute the acceptance-ratio according to eq (\ref{eqn:STAcceptanceRatio})\\
					a $\leftarrow$ r / r'\;
					\vspace{2mm}
					\If {random() < a} {
						u[i] $\leftarrow$ u'\;
						st[i] $\leftarrow$ (s',t')\;
					}
				}
				{
					// apply a standard primary sample space mutation\\
					u' $\leftarrow$ {\bf perturb}(u[i])\;
					\vspace{2mm}
					(path',C') $\leftarrow$ {\bf bptTracePath}(u', st[i])\;
					\vspace{2mm}
					a $\leftarrow$ min(1, C' / C[i])\;
					\vspace{2mm}
					\If {random() < a} {
						u[i] $\leftarrow$ u'\;
						C[i] $\leftarrow$ C'\;
						path[i] $\leftarrow$ path'\;
					}
				}
				
				// accumulate the new sample\\
				{\bf accumulate}(path[i], C[i]);
			}
		}
	}
	\caption{pseudo-code for our CMLT algorithm}
\end{algorithm}

\bibliographystyle{acmsiggraph}
\bibliography{main}

\begin{thebibliography}{\protect\citename{Kaplanyan et~al\mbox{.} }2014}

\bibitem[\protect\citename{Bitterli }2016]{resources16}
{\sc Bitterli, B.}, 2016.
\newblock Rendering resources.
\newblock https://benedikt-bitterli.me/resources/.

\bibitem[\protect\citename{Cline et~al\mbox{.} }2005]{Cline:2005}
{\sc Cline, D., Talbot, J., and Egbert, P.}
\newblock 2005.
\newblock Energy redistribution path tracing.
\newblock In {\em ACM SIGGRAPH 2005 Papers}, ACM, New York, NY, USA, SIGGRAPH
  '05, 1186--1195.

\bibitem[\protect\citename{Geyer }2011]{Geyer:2011}
{\sc Geyer, C.~J.}
\newblock 2011.
\newblock Introduction to {M}arkov {C}hain {M}onte {C}arlo.
\newblock In {\em Handbook of Markov Chain Monte Carlo}. Chapman and Hall CRC,
  ch.~1, 3--47.

\bibitem[\protect\citename{Hachisuka and Jensen }2011]{Hachisuka:2011}
{\sc Hachisuka, T., and Jensen, H.~W.}
\newblock 2011.
\newblock Robust adaptive photon tracing using photon path visibility.
\newblock {\em ACM Trans. Graph. 30}, 5 (Oct.), 114:1--114:11.

\bibitem[\protect\citename{Hachisuka et~al\mbox{.} }2012]{Hachisuka:2012}
{\sc Hachisuka, T., Pantaleoni, J., and Jensen, H.~W.}
\newblock 2012.
\newblock A path space extension for robust light transport simulation.
\newblock {\em ACM Trans. Graph. 31}, 6 (Nov.), 191:1--191:10.

\bibitem[\protect\citename{Hachisuka et~al\mbox{.} }2014]{Hachisuka:2014}
{\sc Hachisuka, T., Kaplanyan, A.~S., and Dachsbacher, C.}
\newblock 2014.
\newblock Multiplexed {M}etropolis light transport.
\newblock {\em ACM Trans. Graph. 33}, 4 (July), 100:1--100:10.

\bibitem[\protect\citename{Hanika et~al\mbox{.} }2015]{Hanika:2015:IHLST}
{\sc Hanika, J., Kaplanyan, A., and Dachsbacher, C.}
\newblock 2015.
\newblock Improved half vector space light transport.
\newblock {\em Computer Graphics Forum (Proceedings of Eurographics Symposium
  on Rendering) 34}, 4 (June), 65--74.

\bibitem[\protect\citename{Heitz and D'Eon }2014]{heitz:hal-2014}
{\sc Heitz, E., and D'Eon, E.}
\newblock 2014.
\newblock {Importance Sampling Microfacet-Based BSDFs using the Distribution of
  Visible Normals}.
\newblock {\em {Computer Graphics Forum} 33}, 4 (July), 103--112.

\bibitem[\protect\citename{Kaplanyan et~al\mbox{.} }2014]{Kaplanyan:2014:HSLT}
{\sc Kaplanyan, A.~S., Hanika, J., and Dachsbacher, C.}
\newblock 2014.
\newblock The natural-constraint representation of the path space for efficient
  light transport simulation.
\newblock {\em ACM Transactions on Graphics (Proc. SIGGRAPH) 33}, 4.

\bibitem[\protect\citename{Kelemen et~al\mbox{.} }2002]{Kelemen:2002}
{\sc Kelemen, C., Szirmay-Kalos, L., Antal, G., and Csonka, F.}
\newblock 2002.
\newblock A simple and robust mutation strategy for the {M}etropolis light
  transport algorithm.
\newblock In {\em Computer Graphics Forum}, 531--540.

\bibitem[\protect\citename{Laine et~al\mbox{.} }2013]{Laine:2013:MCH}
{\sc Laine, S., Karras, T., and Aila, T.}
\newblock 2013.
\newblock Megakernels considered harmful: Wavefront path tracing on gpus.
\newblock In {\em Proceedings of the 5th High-Performance Graphics Conference},
  ACM, New York, NY, USA, HPG '13, 137--143.

\bibitem[\protect\citename{Marinari and Parisi }1992]{Marinari:1992}
{\sc Marinari, E., and Parisi, G.}
\newblock 1992.
\newblock Simulated tempering: A new {M}onte {C}arlo scheme.
\newblock {\em Europhysics Letters}, 19, 451--458.

\bibitem[\protect\citename{Otsu et~al\mbox{.} }2017]{Anon:0462}
{\sc Otsu, H., Kaplanyan, A., Hanika, J., Dachsbacher, C., and Hachisuka, T.}
\newblock 2017.
\newblock {Fusing State Spaces for Markov Chain Monte Carlo Rendering}.
\newblock ACM, SIGGRAPH '17.

\bibitem[\protect\citename{Pantaleoni }2017]{SelfSiggraph:2017}
{\sc Pantaleoni, J.}
\newblock 2017.
\newblock {Charted Metropolis Light Transport}.
\newblock ACM, SIGGRAPH '17.

\bibitem[\protect\citename{Qin et~al\mbox{.} }2015]{Qin:2015:UPG}
{\sc Qin, H., Sun, X., Hou, Q., Guo, B., and Zhou, K.}
\newblock 2015.
\newblock Unbiased photon gathering for light transport simulation.
\newblock {\em ACM Trans. Graph. 34}, 6 (Oct.), 208:1--208:14.

\bibitem[\protect\citename{Swendsen and Wang }1986]{Swendsen:1986}
{\sc Swendsen, R.~H., and Wang, J.-S.}
\newblock 1986.
\newblock Replica {M}onte {C}arlo simulation of spin-glasses.
\newblock {\em Phys. Rev. Lett.}, 57, 2607--2609.

\bibitem[\protect\citename{Tierney }1994]{Tierney:1994}
{\sc Tierney, L.}
\newblock 1994.
\newblock Markov chains for exploring posterior distributions.
\newblock {\em Annals of Statistics 22\/}, 1701--1762.

\bibitem[\protect\citename{Veach and Guibas }1997]{Veach:1997:MLT}
{\sc Veach, E., and Guibas, L.~J.}
\newblock 1997.
\newblock Metropolis light transport.
\newblock In {\em Proceedings of the 24th Annual Conference on Computer
  Graphics and Interactive Techniques}, ACM Press/Addison-Wesley Publishing
  Co., New York, NY, USA, SIGGRAPH '97, 65--76.

\bibitem[\protect\citename{Veach }1997]{Veach:PHD}
{\sc Veach, E.}
\newblock 1997.
\newblock {\em Robust Monte Carlo Methods for Light Transport Simulation}.
\newblock PhD thesis, Stanford University.

\bibitem[\protect\citename{\v{S}ik et~al\mbox{.} }2016]{Sik:2016}
{\sc \v{S}ik, M., Otsu, H., Hachisuka, T., and K\v{r}iv\'{a}nek, J.}
\newblock 2016.
\newblock Robust light transport simulation via {M}etropolised bidirectional
  estimators.
\newblock {\em ACM Trans. Graph. 35}, 6 (Nov.), 245:1--245:12.

\bibitem[\protect\citename{Wilkie et~al\mbox{.} }2014]{Wilkie:2014:HWS}
{\sc Wilkie, A., Nawaz, S., Droske, M., Weidlich, A., and Hanika, J.}
\newblock 2014.
\newblock Hero wavelength spectral sampling.
\newblock In {\em Proceedings of the 25th Eurographics Symposium on Rendering},
  Eurographics Association, Aire-la-Ville, Switzerland, Switzerland, EGSR '14,
  123--131.

\end{thebibliography}

\end{document}